\documentclass[a4paper,10pt]{article}
\usepackage{theorem}

\newtheorem{theorem}{Theorem}[section]

\newtheorem{lemma}[theorem]{Lemma}

\newcount\driver
\newcount\bozza

\font\msytw=msbm10 scaled\magstep1                  
                 \font\indbf=cmbx10 scaled\magstep2

{\count255=\time\divide\count255 by 60 \xdef\hourmin{\number\count255}
        \multiply\count255 by-60\advance\count255 by\time
   \xdef\hourmin{\hourmin:\ifnum\count255<10 0\fi\the\count255}}

\let\a=\alpha \let\b=\beta    \let\g=\gamma     \let\d=\delta     \let\e=\varepsilon
  \let\h=\eta     \let\th=\vartheta      \let\l=\lambda
\let\m=\mu    \let\n=\nu      \let\x=\xi        \let\p=\pi        
\let\s=\sigma \let\t=\tau     \let\f=\varphi       \let\c=\chi
\let\ps=\psi   \let\o=\omega     
\let\G=\Gamma \let\D=\Delta       \let\L=\Lambda    
      \let\F=\Phi

\def\VV{{\cal V}}
\def\WW{{\cal W}}
\def\NN{{\cal N}}\def\BB{{\cal B}}
\def\LL{{\cal L}}\def\QQ{{\cal Q}}
\def\DD{{\cal D}}\def\AA{{\cal A}}\def\SS{{\cal S}}

\def\bfs{{\bf s}}
\def\pp{{\bf p}}\def\qq{{\bf q}}\def\xx{{\bf x}}
   
\def\yy{{\bf y}}\def\kk{{\bf k}}\def\nn{{\bf n}}
\def\zz{{\bf z}}\def\uu{{\bf u}}\def\vv{{\bf v}}\def\ww{{\bf w}}
 
\def\hh{{\bf h}}

       \def\oo{{\underline \omega}}
\def\bfe{{\bf e}} 

          \def\ux{{\underline\xx}}
\def\uk{{\underline \kk}}

           \def\uy{{\underline\yy}}
\def\uz{{\underline \zz}}
          \def\uo{{\underline \o}}\def\oo{{\underline \omega}}

\def\qed{\raise1pt\hbox{\vrule height5pt width5pt depth0pt}}

\def\indic{\hbox{\raise-2pt \hbox{\indbf 1}}}

\def\RRR{\hbox{\msytw R}}

 \def\ZZZ{\hbox{\msytw Z}}

\def\tA{\tilde A}
\def\tit{\tilde t}
\def\tc{\tilde c}
\def\ins#1#2#3{\vbox to0pt{\kern-#2 \hbox{\kern#1 #3}\vss}\nointerlineskip}

\newdimen\xshift \newdimen\xwidth \newdimen\yshift
\newcount\griglia

\def\insertplot#1#2#3#4#5#6{%
\xwidth=#1pt \xshift=\hsize \advance\xshift by-\xwidth \divide\xshift by 2%
\begin{figure}[ht]
\vspace{#2pt}
\hspace{\xshift}
\begin{minipage}{#1pt}
#3
\ifnum\driver=1 \griglia=#6
\ifnum\griglia=1
\openout13=griglia.ps
\write13{gsave .2 setlinewidth}
\write13{0 10 #1 {dup 0 moveto #2 lineto } for}
\write13{0 10 #2 {dup 0 exch moveto #1 exch lineto } for}
\write13{stroke}
\write13{.5 setlinewidth}
\write13{0 50 #1 {dup 0 moveto #2 lineto } for}
\write13{0 50 #2 {dup 0 exch moveto #1 exch lineto } for}
\write13{stroke grestore}
\closeout13
\includegraphics{griglia.ps}
\fi
\includegraphics{#4.ps}\fi%
\ifnum\driver=2 \fi
\end{minipage}
\caption{#5}
\end{figure}
}

\def\be{\begin{equation}}
\def\ee{\end{equation}}
\def\bea{\begin{eqnarray}}\def\eea{\end{eqnarray}}
\def\bean{\begin{eqnarray*}}\def\eean{\end{eqnarray*}}
\def\bfr{\begin{flushright}}\def\efr{\end{flushright}}
\def\bc{\begin{center}}\def\ec{\end{center}}
\def\bal{\begin{align}}\def\eal{\end{align}}
\def\ba#1{\begin{array}{#1}} \def\ea{\end{array}}
\def\bd{\begin{description}}\def\ed{\end{description}}
\def\bv{\begin{verbatim}}\def\ev{\end{verbatim}}
\def\nn{\nonumber}
\def\Halmos{\hfill\vrule height10pt width4pt depth2pt \par\hbox to \hsize{}}
\def\pref#1{(\ref{#1})}
 
\def\virg{\quad,\quad}

\let\a=\alpha \let\b=\beta    \let\g=\gamma     \let\d=\delta     \let\e=\varepsilon
  \let\h=\eta     \let\th=\vartheta      \let\l=\lambda
\let\m=\mu    \let\n=\nu      \let\x=\xi        \let\p=\pi        
\let\s=\sigma \let\t=\tau     \let\f=\varphi       \let\c=\chi
\let\ps=\psi   \let\o=\omega     
\let\G=\Gamma \let\D=\Delta       \let\L=\Lambda    
      \let\F=\Phi

\def\VV{{\cal V}}
\def\WW{{\cal W}}
\def\NN{{\cal N}}\def\BB{{\cal B}}
\def\LL{{\cal L}}\def\QQ{{\cal Q}}
\def\DD{{\cal D}}\def\AA{{\cal A}}\def\SS{{\cal S}}

\def\pp{{\bf p}}\def\qq{{\bf q}}\def\xx{{\bf x}}
   
\def\yy{{\bf y}}\def\kk{{\bf k}}\def\nn{{\bf n}}
\def\zz{{\bf z}}\def\uu{{\bf u}}\def\vv{{\bf v}}\def\ww{{\bf w}}

       \def\oo{{\underline \omega}}
\def\ee{{\underline \varepsilon}}  
          
          \def\ux{{\underline\xx}}
\def\uk{{\underline \kk}}

           \def\uy{{\underline\yy}}
\def\uz{{\underline \zz}}
          \def\uo{{\underline \o}}

\def\RRR{\hbox{\msytw R}}

        \def\ZZZ{\hbox{\msytw Z}}

\let\dpr=\partial

\let\==\equiv

\let\io=\infty
\let\0=\noindent

\def\*{{\hfill\break\null\hfill\break}}

\def\eg{\hbox{\it e.g.\ }}

\def\tilde#1{{\widetilde #1}}

\def\lft{\left}
\def\rgt{\right}

\def\la{{\langle}}
\def\ra{{\rangle}}

\def\tende#1{\,\vtop{\ialign{##\crcr\rightarrowfill\crcr
             \noalign{\kern-1pt\nointerlineskip}
             \hskip3.pt${\scriptstyle #1}$\hskip3.pt\crcr}}\,}
\def\otto{\,{\kern-1.truept\leftarrow\kern-5.truept\to\kern-1.truept}\,}

\def\defi{{\buildrel \;def\; \over =}}

\def\wh#1{\widehat{#1}}
\def\hat#1{\wh{#1}}
\def\sqt[#1]#2{\root #1\of {#2}}

\def\ha{{\widehat \a}}\def\hb{{\widehat \b}}
\def\hv{{\widehat v}}
\def\hW{{\widehat W}}
\def\hh{{\widehat \h}}\def\hu{{\widehat u}}
\def\hK{{\widehat K}} \def\hW{{\widehat W}}\def\hU{{\widehat U}}
\def\hp{{\widehat \ps}}  \def\hF{{\widehat F}}

\def\hc{{\hat \c}}

\def\hJ{{\widehat \jmath}}
\def\hJ{{\widehat J}}
\def\hg{{\widehat g}}

\def\hA{{\widehat A}}

\def\hG{{\widehat G}}
\def\hS{{\widehat S}}

\def\VV{{\cal V}}
\def\WW{{\cal W}}
\def\NN{{\cal N}}\def\BB{{\cal B}}
\def\LL{{\cal L}}\def\QQ{{\cal Q}}
\def\DD{{\cal D}}\def\AA{{\cal A}}\def\SS{{\cal S}}
\def\AAA{{\cal A}}

\def\T#1{{#1_{\kern-3pt\lower7pt\hbox{$\widetilde{}$}}\kern3pt}}
\def\VVV#1{{\underline #1}_{\kern-3pt
\lower7pt\hbox{$\widetilde{}$}}\kern3pt\,}
\def\W#1{#1_{\kern-3pt\lower7.5pt\hbox{$\widetilde{}$}}\kern2pt\,}

\def\indica{\leaders \hbox to 0.5cm{\hss.\hss}\hfill}
\def\guida{\leaders\hbox to 1em{\hss.\hss}\hfill}
\mathchardef\oo= "0521

\def\pp{{\bf p}}\def\qq{{\bf q}}\def\xx{{\bf x}}
\def\yy{{\bf y}}\def\kk{{\bf k}}\def\nn{{\bf n}}
\def\zz{{\bf z}}\def\uu{{\bf u}}\def\vv{{\bf v}}

\def\oo{{\underline \omega}}

\def\qed{\raise1pt\hbox{\vrule height5pt width5pt depth0pt}}

\def\indic{\hbox{\raise-2pt \hbox{\indbf 1}}}

\def\RRR{\hbox{\msytw R}}

 \def\ZZZ{\hbox{\msytw Z}}

\def\ins#1#2#3{\vbox to0pt{\kern-#2 \hbox{\kern#1 #3}\vss}\nointerlineskip}

\newdimen\xshift \newdimen\xwidth \newdimen\yshift
\newcount\griglia

\def\insertplot#1#2#3#4#5#6{%
\xwidth=#1pt \xshift=\hsize \advance\xshift by-\xwidth \divide\xshift by 2%
\begin{figure}[ht]
\vspace{#2pt} \hspace{\xshift}
\begin{minipage}{#1pt}
#3 \ifnum\driver=1 \griglia=#6
\ifnum\griglia=1 \openout13=griglia.ps \write13{gsave .2
setlinewidth} \write13{0 10 #1 {dup 0 moveto #2 lineto } for}
\write13{0 10 #2 {dup 0 exch moveto #1 exch lineto } for}
\write13{stroke} \write13{.5 setlinewidth} \write13{0 50 #1 {dup 0
moveto #2 lineto } for} \write13{0 50 #2 {dup 0 exch moveto #1
exch lineto } for} \write13{stroke grestore} \closeout13
\includegraphics{griglia.ps} \fi
\includegraphics{#4.ps}\fi%
\ifnum\driver=2 \fi
\end{minipage}
\caption{#5}
\end{figure}
}

\newdimen\shift \shift=-1.5truecm
\def\lb#1{%
\ifnum\bozza=1
\label{#1}\rlap{\hbox{\hskip\shift$\scriptstyle#1$}}
\else\label{#1} \fi}

\def\be{\begin{equation}}
\def\ee{\end{equation}}
\def\bea{\begin{eqnarray}}\def\eea{\end{eqnarray}}
\def\bean{\begin{eqnarray*}}\def\eean{\end{eqnarray*}}
\def\bfr{\begin{flushright}}\def\efr{\end{flushright}}
\def\bc{\begin{center}}\def\ec{\end{center}}
\def\bal{\begin{align}}\def\eal{\end{align}}
\def\ba#1{\begin{array}{#1}} \def\ea{\end{array}}
\def\bd{\begin{description}}\def\ed{\end{description}}
\def\bv{\begin{verbatim}}\def\ev{\end{verbatim}}
\def\nn{\nonumber}
\def\Halmos{\hfill\vrule height10pt width4pt depth2pt \par\hbox to \hsize{}}
\def\pref#1{(\ref{#1})}
 
\def\virg{\quad,\quad}

\title{Extended scaling relations for planar lattice models}

\author{G. Benfatto$^1$ \and P. Falco$^2$ \and  V. Mastropietro$^1$\\
\\
{\small $^{1}$ Dipartimento di Matematica, Universit\`a di Roma ``Tor Vergata''}\\ 
{\small via della Ricerca Scientifica, I-00133, Roma}\\
\\
{\small $^{2}$   Mathematics Department, University of British Columbia,}\\
{\small Vancouver, BC Canada, V6T 1Z2}}
\date{}
\begin{document}
\maketitle

\begin{abstract} It is widely believed that the critical properties
of several planar lattice models, like the Eight Vertex or the
Ashkin-Teller models, are well described by
an effective Quantum Field Theory obtained as formal scaling
limit.  On the basis of this
assumption several extended scaling relations among their
indices were conjectured. We prove the validity of some of them, among
which the ones by Kadanoff, \cite{[K]}, and
by Luther and Peschel, \cite{[LP]}.
\end{abstract}

\section{Introduction and main results}

Integrable models in statistical mechanics, like the Ising or the Eight vertex
(8V) models in two dimensions, provide conceptual laboratories for the
understanding of phase transitions. Integrability is however a rather delicate
property requiring very special features, and it is usually lost in more
realistic models.

The principle of {\it universality}, phenomenologically quite well verified,
says that the singularities for second order phase transitions should be
insensitive to the specific details of the model, provided that symmetry and
some form of locality are retained. From the theoretical side, a mathematical
justification of universality in planar lattice models is rather complex to
provide. Only very recently Pinson and Spencer established, see
\cite{[Sp],[PS]}, a form of universality for the 2D Ising model; they added to
the Ising Hamiltonian a perturbation breaking the integrability and showed that
the indices they can compute were {\it exactly the same} as the Ising model
ones.

While the critical indices of the Ising model are expressed by {\it pure
numbers}, there are other lattice models in which some of the critical
exponents vary continuously with the parameters appearing in the Hamiltonian. A
celebrated example is provided by the Eight vertex model, solved by Baxter in
[2]; even if it can be mapped in two Ising models coupled by a quartic
interaction, its critical indices are different from the Ising ones.

Several authors, starting from Kadanoff and collaborators \cite{[K],[KB],[KW]}
and Luther and Peschel \cite{[LP]} , have argued that many models, like the
{\it Askhin-Teller} (AT) model and several others, belongs to the class of
universality of the 8V model. The notion of universality in this case is much
more subtle; it does not mean that the indices are the same for all the models
in the same class (on the contrary, the indices depend on all details of the
Hamiltonian), but that there are {\it scaling relations} between them, such
that all indices can be expressed in terms of any one of them.

The notion of universality for models with continuously varying
indices has been deeply investigated over the years, see for
instance \cite{[KW],[N],[PB],[ZZ]}; it has been pointed out that such
models are well described in the scaling limit by an effective
Quantum Field Theory, and on the basis of this assumption several
extended scaling relations between their indices were derived.
While the assumption of continuum scaling limit description of
planar lattice models is very powerful, it is well known that a
mathematical justification of it is very difficult, see \eg
\cite{[Sm]}.

The aim of this paper is to provide a mathematical proof of some of the exact
scaling relations derived in the literature for planar lattice models. We will
focus mainly on the 8V and AT models, but, as we will explain after the main
theorem below, our result can be extend to several other models.

We start from the well known (see \cite{[Ba]}) Ising formulation
of the 8V and the AT models. Let
$\L$ be a square subset of $\ZZZ^2$ of side $L$; if $\xx=(x_0,x)\in \L$ and
$\bfe_0=(1,0)$, $\bfe_1=(0,1)$, we consider two independent configurations of
spins, $\{\s_\xx=\pm1\}_{\xx\in \L }$ and $\{\s'_\xx=\pm1\}_{\xx\in \L}$ and
the Hamiltonian
\bea\lb{111} &&H(\s,\s')=H_J(\s) + H_{J'}(\s')- J_4 V(\s,\s')\;,
\eea
where $J>0$ and $J'>0$ are two parameters, $H_J$ is the (ferromagnetic) Ising
Hamiltonian in the lattice $\L$,
\bea
H_J(\s)=- J\sum_{j=0,1}\sum_{\xx\in\L} \s_{\xx}\s_{\xx+\bfe_j}\;,
\eea
$V$ is the quartic interaction and $-J_4$ is the coupling. In the the AT model,
$J$ and $J'$ can be different (that case is called {\it anisotropic}) and
$V=V_{AT}$, with
\bea V_{AT}(\s,\s')=
\sum_{j=0,1}\sum_{\xx\in\L}\s_\xx\s_{\xx+\bfe_j} \s'_\xx\s'_{\xx+\bfe_j}\;.
\eea
In the 8V model $J=J'$ and $V=V_{8V}$, with
\bea V_{8V}(\s,\s')=
\sum_{j=0,1}\sum_{\xx\in\L}\s_{\xx+j(\bfe_0+\bfe_1)}\s_{\xx+\bfe_0}
\s'_{\xx+j(\bfe_0+\bfe_1)}\s'_{\xx+\bfe_1}\;.
\eea
\insertplot{240}{70}
{}%
{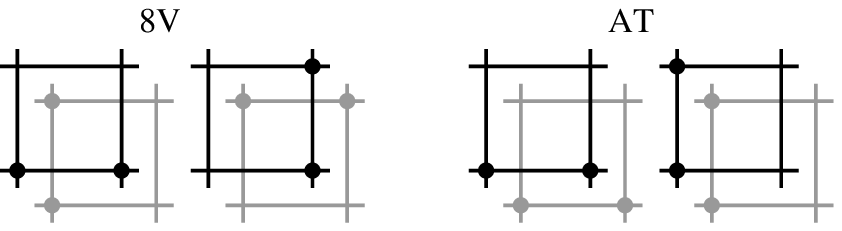}{\lb{s5}: The quartic interaction in the 8V and in the AT case. The
gray and the black square are the same square of the lattice. }{0}
In this paper we will focus our attention on two observables,
\bea
&&O^\e_\xx=\sum_{j=0,1}\s_{\xx}\s_{\xx+\bfe_j}+
\e\sum_{j=0,1}\s'_{\xx}\s'_{\xx+\bfe_j}\virg \e=\pm\;,
\eea
and their truncated correlations in the {\it thermodynamic limit}
\be\lb{corr} G^\e(\xx-\yy)= \lim_{\L\to \io}\la O^\e_\xx O^\e_\yy\ra_{\L} -\la
O^\e_\xx\ra_{\L}\la O^\e_\yy\ra_{\L}\virg \e=\pm\;,
\ee
where $\la\cdot\ra_\L$ is the average over all configurations of the spins with
statistical weight  $e^{-\b H(\s,\s')}$. In the AT model, $\la O^+_\xx\ra$ is
called the {\it energy}, while $\la O^-_{\xx}\ra$ is called the  {\it
crossover}; in the 8V model is the opposite, see \eg \cite{[N]}.

Despite their similarity, an exact solution exists for the 8V model but {\it
not} for the AT model. In recent times the methods of {\it constructive}
fermionic Renormalization (see \eg \cite{[M4]} for an updated introduction) has
been applied to such models, using the well known representation of such
correlations in terms of {\it Grassmann integrals}, see \eg \cite{[Sa]}. It was
proved in \cite{[M0],[M1]} that both the 8V and the isotropic AT systems have a
nonzero {\it critical temperature}, $T_c$, such that, if $T\not= T_c$,
$G^\e(\xx-\yy)$ decays faster than any power of $\x |\xx-\yy|$, with
\be
\x \sim  C\,|T-T_c|^{\a} \;, \hbox{as\ } T\to T_c \;.
\ee
Moreover, at criticality, there are two constants $C_\e$, $\e=\pm$, such that
\be\lb{xpm}
G^\e(\xx-\yy)\sim {C_\e\over |\xx-\yy|^{2 x_\e}} \;, \hbox{as\ }
|\xx-\yy|\to\io\;,
\ee
where $x_\pm$ are critical indices expressed by {\it convergent series} in
$J_4$.
The analysis in \cite{[M1]} allows to compute the indices $\a,x_\pm$ with
arbitrary precision (by an explicit computation of the lowest orders and a
rigorous bound on the rest); the complexity of such expansions makes however
essentially impossible to see directly from them the extended scaling
relations.

In the case of the anisotropic AT model, it was proven in  \cite{[GM1]}
that there are two critical temperatures,
$T_{1,c}$ and $T_{2,c}$, and the corresponding critical indices are
the same as those of the Ising model. However as
$J-J'\to 0$:
\be \lb{tr}
|T_{1,c}-T_{2,c}|\sim |J-J'|^{x_T}\;,
\ee
with a {\it transition index}, $x_T$, different form $1$ if $J_4\neq 0$.

In this paper we will prove the following Theorem.

\begin{theorem}\lb{thm1}
If the coupling is small enough, the critical indices of the 8V or AT verify
\be\lb{2}
x_-={1\over x_+}\;, \ee
\be\lb{2a}
\a={1\over 2-x_+}\;;
\ee
and, in the case of the anisotropic AT model,
\be\lb{3}
x_T={2-x_+\over 2-x_-}\;.
\ee
Moreover, if $-J_4^{AT}$ and $-J_4^{8V}$ denote the coupling in the two models,
there exists a choice of $J_4^{AT}$ as function of $J_4^{8V}$ such that the
above critical indices coincide.
\end{theorem}

{\bf Remarks}
\begin{enumerate}
\item Equation \pref{2} is the {\it extended scaling law} first conjectured
    by Kadanoff for the AT and 8V models, mainly on the basis of numerical
    evidence (see eq.(13b) and (15b) of \cite{[K]}). The scaling relation
    \pref{3} was never conjectured before. Note that all the critical
    indices we consider can be expressed as simple functions of one of
    them, in agreement with the general belief.

\item A similar theorem can be proved for a number of other models
in the same class of universality. An example is provided by the
$XYZ$ model, describing the nearest-neighbor interaction of
quantum spins on a chain with couplings $J_1,J_2,J_3$. In
\cite{[BM1]}, by a rigorous Renormalization Group analysis valid
for small values of $J_3$, it was possible to write two critical
indices as a convergent series in $J_3$; there were the index
$1+\h_1$, appearing in the oscillating part of the spin-spin
correlation along the $z$ direction (see (1.20) of \cite{[BM1]}),
and $1+\h_2$, the index appearing in the decay rate (see (1.19) of
\cite{[BM1]}). In such a case the analogue of the second of (1.10)
can be written as
\be
1+\h_2={1\over 2-2(1+\h_1)^{-1}}
\ee
The above relation for the $XYZ$ indices has been conjectured
by Luther and Peschel in
\cite{[LP]} (see eq.(16) and table I of that paper).

\item Our results could be easily extended to any Hamiltonian of the form
    \pref{111}, if the quartic interactions verifies some symmetry
    conditions, listed in App. O of \cite{[M1]}.

\item Several other relations are conjectured in the literature,
concerning critical indices which are much more difficult to study
with our methods, like the indices of the polarization
correlations. New ideas seems to be required to treat such cases.
\end{enumerate}

The paper is organized in the following way. In \S\ref{sec2} we summarize the
analysis given in [18,19], in which the correlations of the AT or 8V models are
written in terms of Grassmann integrals and are analyzed using constructive
Renormalization Group methods. The outcome of such analysis is that the
critical indices $x_+$, $x_-$, $\a$ and $x_T$ can be written, in the small
coupling region, as {\it model independent} convergent series of a single
parameter, $\l_{-\io}$, the asymptotic limit of the effective coupling on large
scale. Note that $\l_{-\io}$ is in turn a convergent series (that does depend
on all the details of the lattice model) of the coupling $J_4$.  Such
expansions allow in principle to compute the indices with arbitrary precision,
but this is not needed to prove \pref{2a} and \pref{3}, which simply follow
from dimensional arguments. On the contrary,  dimensional arguments are not
sufficient to prove \pref{2}; and  it is apparently impossible to check it
directly in terms of the series representing $x_+$ and $x_-$, as functions of
$\l_{-\io}$.

In \S\ref{sec3} we show that such indices are {\it equal} to the indices of the
Quantum Field Theory coinciding with the formal scaling limit of the spin
models, provided the {\it bare parameters} of such a theory are chosen properly
as suitable functions of the parameters of the 8V or AT models; such functions
are expressed in terms of convergent expansions depending on all details of the
spin models. On the other hand, the QFT verifies extra quantum symmetries with
respect to the original spin Hamiltonian \pref{111}, implying a set of {\it
Ward Identities} and closed equations allowing to get simple exact expressions
for the critical indices in terms of the coupling of the QFT; \pref{2} follows
from such expressions.

\section{RG analysis of spin models}\lb{sec2}

\subsection{Fermionic representation  of the partition function}
We begin with considering  the partition function of
the Ising model with a quadratic interaction, external sources $A_{j,\xx}$, and
periodic conditions at the boundary of $\L$:
\be\lb{pf1}
Z(I)=\sum_{\s}\exp\Big[
\sum_{j=0,1\atop \xx\in \L}I_{j,\xx}\s_\xx\s_{\xx+\bfe_j}\Big]
\ee
where $I_{j,\xx}=A_{j,\xx}+\b J$. The purpose of adding the external source is
twofold: by taking derivatives w.r.t. $A$, either we can write the partition
function for \pref{111} in terms of two non-interacting Ising models, or we can
generate the correlations of the quadratic observables.

Indeed, since $\s_\xx, \s'_\xx=\pm1$,
$$\exp\big(\a\s_\xx \s_{\xx+\bfe_j}\s'_\yy \s'_{\yy+\bfe_{j'}}\big)
=\cosh(\a)+\s_\xx \s_{\xx+\bfe_j}\s'_\yy \s'_{\yy+\bfe_{j'}}\sinh(\a)\;,$$
so that the partition function of the two models with external fields is given
by:
\bea\lb{int}
&&Z(J_4,I,I') = \lft[\cosh(\b J_4)\rgt]^{2|\L|}\cdot\nn\\
&&\cdot \prod_{j=0,1\atop\xx\in\L}\lft[1+\tanh(\b J_4){\dpr^2\over \dpr
\tA_{j,\xx} \tA'_{j,\xx}}\rgt] Z(I)Z(I')\;,
\eea
where $I'_{j,\xx}=A'_{j,\xx}+\b J'$; and,
in the AT case, $\tA_{j,\xx}=A_{j,\xx}$ and $\tA'_{j,\xx}=A'_{j,\xx}$,
while, in the 8V case, $\tilde A_{0,\xx}=A_{0,\xx}$, $\tilde
A'_{0,\xx}=A'_{1,\xx}$, $\tilde A_{1,\xx}=A_{1,\xx+\bfe_0}$, $\tilde
A'_{1,\xx}=A'_{0,\xx+\bfe_1}$.

For $Z(I)$, the partition function of the Ising model with periodic boundary
condition, a fermionic representation is known since a long time, see
\cite{[Sa]}.

The result is the following. Let $\g=(\e_0,\e_1)$, with $\e_0,\e_1=\pm$ and let
$\{H_\xx$, $\bar H_\xx$, $V_\xx$, $\bar V_\xx\}_{\xx\in \L}$ be a family of
Grassmann variables verifying the $\g$-boundary conditions, namely
\bea &&\bar H_{\xx+(L,0)}=\e_0\bar H_{\xx} \virg
\bar H_{\xx+(0,L)}=\e_1\bar H_{\xx}\nn\;,\\
\lb{c31} && H_{\xx+(L,0)}=\e_0 H_{\xx} \virg H_{\xx+(0,L)}=\e_1 H_{\xx}\;,
\eea
and similar relations for $V,\bar V$ (we are skipping the $\g$ dependence in
$H$'s and $V$'s). Then  we consider the {\it Grassmann functional integral}
\be \lb{2.11}
Z_\g= \int\! dH dV\; e^{S(t)}\;,
\ee
where the action $S(t)$ is the following function of the parameters
$t=\{t_{j,\xx}\}_{\xx\in \L\atop j=0,1}$ and of the Grassmann variables with
$\g-$boundary condition:
\bea \lb{16}
&& S(t)= \sum_{\xx\in\L} \Big[t_{0,\xx}\bar H_{\xx} H_{\xx+\bfe_0}+
t_{1,\xx}\bar V_\xx V_{\xx+\bfe_1}\Big]+\\
&&+\sum_{\xx\in\L}\Big[ \bar H_{\xx} H_{\xx}+ \bar V_{\xx} V_{\xx}+ \bar
V_{\xx} \bar H_{\xx}+  V_{\xx} H_{\xx}+ V_{\xx} \bar H_{\xx}+ H_{\xx} \bar
V_{\xx}\Big]\nn\;.
\eea
Choosing $t_{j,\xx}=\tanh I_{j,\xx}$, and for $c_{j,\xx}=\cosh I_{j,\xx}$, the
partition function \pref{pf1} can be written in the following way:
\bea\lb{17}
Z(I) = (-1)^{|\L|} 2^{|\L|}\lft(\prod_{j,\xx}c_{j,x}\rgt) \sum_{\g}
{(-1)^{\d_\g}\over 2}Z_{\g}
\eea
where $\d_\g=1$ for $\g=(+,+)$, and $\d_\g=0$ otherwise.

By \pref{int},  $Z( J_4,I,I')$ can be written by doubling the above
representation and explicitly taking the derivatives w.r.t. $\tA_{j,\xx}$ and
$\tA'_{j,\xx}$. After some trivial algebra, we get the following result.

Let us call $\tit_{j,\xx}$, $\tc_{j,\xx}$ the expressions obtained from
$t_{j,\xx}$, $c_{j,\xx}$ by substituting $A_{j,\xx}$ with $\tA_{j,\xx}$; in a
similar way we define $\tit'_{j,\xx}$, $\tc'_{j,\xx}$. Let us now define:
\bea
&& f_{j,\xx} = 1+\tanh(\b J_4) \tit_{j,\xx} \tit'_{j,\xx}\;,\nn\\
&& g_{j,\xx}={\tit'_{j,\xx}\over (\tc_{j,\xx})^2 } {\tanh(\b J_4)\over
f_{j,\xx}} \virg g'_{j,\xx}= {\tit_{j,\xx}\over (\tc'_{j,\xx})^2 } {\tanh(\b
J_4)\over f_{j,\xx}} \;,\nn\\
&& h_{j,\xx}= {1\over (\tc'_{j,\xx})^2 (\tc_{j,\xx})^2} {\tanh(\b J_4)\over
 f_{j,\xx}} - g_{j,\xx} g'_{j,\xx}\;.
\eea
Then we can write the partition function of the interacting models as
\bea \lb{2.10}
&& Z( J_4,I,I') = 4^{|\L|} \lft[\cosh (\b J_4)\rgt]^{2|\L|}
\lft(\prod_{j,\xx}f_{j,\xx} c_{j,\xx} c'_{j,\xx}\rgt)\cdot\nn\\
&&\cdot \sum_{\g,\g'}{(-1)^{\d_{\g}+\d_{\g'}}\over 4} Z_{\g,\g'}( J_4)\;,
\eea
where $Z_{\g,\g'}(J_4)$ is the Grassmannian functional integral
\bea \lb{2.111a}
Z_{\g,\g'}( J_4)=\int\! dH dV dH' dV'\; e^{\tilde S(\tit+g)+ \tilde
S'(\tit'+g')+V(h)}\;,
\eea
with boundary conditions $\g=(\e_0,\e_1)$ and $\g'=(\e'_0,\e'_1)$ on the
variables $H$, $V$ and $H'$, $V'$, respectively. Moreover $\tilde S(t)$ and
$\tilde S'(t)$ have a definition which depends on the model. $\tilde S(t)$ is
equal to $S(t)$ in the AT model, while, in the 8V model, it is the function
which is obtained from $S(t)$, by substituting, in the first line of \pref{16},
$\bar V_\xx V_{\xx+\bfe_1}$ with $\bar V_{\xx+\bfe_0} V_{\xx+\bfe_0+\bfe_1}$.
$\tilde S'(t)$, in the AT case, is obtained from $S(t)$, by simply replacing
$H, V$ with $H', V'$, while, in the 8V case, we also have to substitute $\bar
H'_\xx H'_{\xx+\bfe_0}$ with $\bar V'_\xx V'_{\xx+\bfe_1}$ and $\bar V'_\xx
V'_{\xx+\bfe_1}$  with $\bar H'_{\xx+\bfe_1} H'_{\xx+\bfe_1+\bfe_0}$. Finally,
$V(h)$ is a quartic interaction that, in the AT case, is given by
\be V_{AT}(h) =\sum_{\xx\in\L}
\lft[h_{0,\xx}\bar H_{\xx} H_{\xx+\bfe_0} \bar H'_{\xx} H'_{\xx+\bfe_0}
+h_{1,\xx}\bar V_{\xx} V_{\xx+\bfe_1} \bar V'_{\xx} V'_{\xx+\bfe_1}\rgt]\;,
\ee
while, in the 8V case, is given by
\be V_{8V}(h) =\sum_{\xx\in\L}
\lft[h_{0,\xx}\bar H_{\xx} H_{\xx+\bfe_0} \bar V'_{\xx} V'_{\xx+\bfe_1} +
h_{1,\xx}\bar V_{\xx+\bfe_0} V_{\xx+\bfe_0+\bfe_1} \bar H'_{\xx+\bfe_1}
H'_{\xx+\bfe_1+\bfe_0}\rgt]\;.
\ee

We remark that
\be
g_{j,\xx},\  g'_{j,\xx},\ h_{j,\xx}={\rm O}(\b J_4)\;.
\ee

\subsection{Fermionic representation of the correlations}

The truncated correlations of the quadratic observables are obtained by taking
two derivatives of $\ln Z( J_4,I,I')$ w.r.t. the external sources in two
different points, and putting such external sources to zero. The
addends $2|\L|\ln [2\cosh(\b J_4)]$ and $\sum_{j,\xx} (\ln f_{j,\xx}+ \ln
c_{j,\xx} + \ln c'_{j,\xx})$ do not contribute when we take two derivatives in
the $A$ variables of two different points. Moreover, it has been proved in
\cite{[M1]} that all and 16 partition functions $Z_{\g,\g'}$ have the same
thermodynamic limit; hence, from now on we will substitute them with the same
one, that with $\g=\g'=(-,-)$. If we define $\dpr^\e_{j,\xx} = \dpr/\dpr
A_{j,\xx} +\e \dpr/\dpr A'_{j,\xx}$, we get:
\be
\la O^\e_\xx;O^\e_\yy\ra_\L^T= \lft.\sum_{i,j} \dpr^\e_{i,\xx} \dpr^\e_{j,\yy}
\ln Z_{\g,\g}\rgt|_{A\=0} = \lft. {\dpr^2\ln \bar Z (\bar A)\over \dpr \bar
A^\e_{\xx}\dpr \bar A^\e_{\yy}} \rgt|_{\bar A\=0}
\ee
where
\be\lb{25}
\bar Z(\bar A)=\int\! dH dV dH' dV'\; e^{S(s)+ S(s')+ 2\l V + B(\bar A)}\;,
\ee
$s$, $s'$ and $h$ are $j,\xx-$independent parameters, defined as
\bea\lb{stg}
&&s=\lft. t_{j,\xx}+g_{j,\xx}\rgt|_{A\=0}=\tanh(\b J) + {\rm O}(\b J_4)
\cr\cr
&&
s'=\lft.
t'_{j,\xx}+g'_{j,\xx}\rgt|_{A\=0}=\tanh(\b J') + {\rm O}(\b J_4)
\cr\cr
&&2\l=\lft. h_{j,\xx}\rgt|_{A\=0}={\rm O}(\b J_4)\;;
\eea
$B(\bar A)$ is an interaction with external sources $\bar A^\e_\xx$, given, in
the AT case, by
\bea
B(\bar A)&=&\sum_{\xx\in \L\atop \e=\pm}
 \bar A^\e_\xx\lft[q_\e\lft(\bar H_{\xx}H_{\xx+\bfe_0}+\bar V_{\xx}V_{\xx+\bfe_1}\rgt)
+ q'_\e \lft(\bar H'_{\xx}H'_{\xx+\bfe_0}+\bar
V'_{\xx}V'_{\xx+\bfe_1}\rgt)\rgt]+\nn\\
&+&\sum_{\xx\in \L\atop \e=\pm} \bar A^\e_{\xx} p_\e\lft(\bar
H_{\xx}H_{\xx+\bfe_0} \bar H'_{\xx}H'_{\xx+\bfe_0}+ \bar V_{\xx} V_{\xx+\bfe_1}
\bar V'_{\xx}V'_{\xx+\bfe_1}\rgt)\;,
\eea
while, in the 8V case, it is given by
\bea
&&B(\bar A) = \sum_{\xx\in \L, \e=\pm} \bar A^\e_\xx \lft[q_\e\lft(\bar
H_{\xx}H_{\xx+\bfe_0}+\bar V_{\xx+\bfe_0} V_{\xx+\bfe_0+\bfe_1}\rgt) +\right.\nn\\
&&\left. +q'_\e \lft(\bar H'_{\xx+\bfe_1}H'_{\xx+\bfe_1+\bfe_0}+\bar
V'_{\xx}V'_{\xx+\bfe_1}\rgt)\rgt]+\\
&&+\sum_{\xx\in \L\atop \e=\pm} \bar A^\e_{\xx} p_\e\lft(\bar
H_{\xx}H_{\xx+\bfe_0} \bar V'_{\xx} V'_{\xx+\bfe_1}+ \bar V_{\xx+\bfe_0}
V_{\xx+\bfe_0+\bfe_1} \bar H'_{\xx+\bfe_1} H'_{\xx+\bfe_1+\bfe_0}\rgt)\;;\nn
\eea
finally, $q_\e$, $q'_\e$ and $p_\e$ are given by the $j,\xx-$independent
parameters
\bea
q_\e&=&\lft. \sum_i \left( {\dpr \over\dpr A_{j,\xx} } +\e {\dpr \over\dpr
A'_{j,\xx} }\right) (\tit_{i,\xx}+g_{i,\xx}) \rgt|_{A\=0} \virg
q'_\e=\{\tit,g \rightarrow \tit',g'\}\;,\nn\\
p_\e&=&\lft. \sum_i \left( {\dpr h_{i,\xx} \over\dpr A_{j,\xx} } + \e {\dpr
h_{j,\xx} \over\dpr A'_{j,\xx} } \right) \rgt|_{A\=0}\;.
\eea
Note that $q_\e=1-\tanh(\b J) + O(\b J_4)$, $q'_\e=\e[1-\tanh(\b J')] + O(\b
J_4)$ and $p_\e=O(\b J_4)$.

\subsection{Dirac and Majorana fermions}
In order to make more evident the analogy of the above functional integral with
the action of a fermionic (Euclidean) Quantum Field Model, it is convenient to
make a change of variables in the Grassmann algebra. This change of variables
is the analogous in the euclidean theories of the transformation from {\it
Dirac fermions} to {\it Majorana fermions} in real time QFT.

The new Grassmannian variables will be denoted by $\psi_\xx$, $\bar\psi_\xx$,
$\c_\xx$ and $\bar\c_\xx$ and are related to the old ones by the equations:
\bea \lb{2.12}
&& \bar H_\xx+i H_\xx= e^{i{\pi\over 4}}\lft(\psi_\xx - \chi_\xx\rgt) \virg
\bar V_\xx+i V_\xx= \psi_\xx + \chi_\xx\;,\nn\\
&& \bar H_\xx -i H_\xx= e^{-i{\pi\over 4}} \lft(\bar\psi_\xx -
\bar\chi_\xx\rgt) \virg \bar V_\xx-i V_\xx= \bar\psi_\xx + \bar\chi_\xx\;.
\eea
A similar transformation is done for the primed variables. After a
straightforward computation, we see that the action \pref{16}, calculated at
$t_{j,\xx}=s$, $\forall j,\xx$, can be written in terms of the Majorana fields
as
\be
S(s)=A(\ps,m_s)+A(\c,M_s)+Q(\ps,\c)\;,
\ee
where $m_s=1-\sqrt{2}+s$, $M_s=1+\sqrt{2}+s$ and, if we define $\dpr^i\psi_\xx
= \psi_{\xx+\bfe_i}-\psi_\xx$,
\bea
A(\ps,m)&=&{s\over 4}\sum_{\xx\in \L} \lft[\ps_\xx \lft(\dpr^0-i\dpr^1\rgt)
\ps_\xx + {\rm c.c.}\rgt] - i m \sum_{\xx\in \L} \bar\ps_\xx
\ps_\xx+\nn\\
&+&{s\over 4}\sum_{\xx\in \L} \lft[\bar\ps_\xx \lft(-i\dpr^0
-i\dpr^1\rgt)\ps_\xx +{\rm c.c.}\rgt]\;,
\eea
\bea
Q(\ps,\c)=&-& {s\over 4}\sum_{\xx\in \L} \lft[\ps_\xx \lft(\dpr^0+i\dpr^1\rgt)
\c_\xx +\Big\{\ps\leftrightarrow \c\Big\}+{\rm c.c.}\rgt]-\nn\\
&-&{s\over 4}\sum_{\xx\in \L} \lft[\bar\c_\xx\lft(-i\dpr^0+i\dpr^1\rgt) \ps_\xx
+\Big\{\ps\leftrightarrow \c\Big\}+{\rm c.c.}\rgt]\;,
\eea
where, in agreement with \pref{2.12}, we are calling complex conjugation (c.c.)
the operation on the Grassmann algebra which amounts to exchange $\psi_\xx$
with $\bar\psi_\xx$, $\c_\xx$ with $\bar\c_\xx$ and $i$ with $-i$.

The quartic interaction of the AT model becomes:
\bea
&&V_{AT} =-\l\sum_{\xx\in \L}\lft[\bar\ps_\xx \ps_\xx \bar\ps'_\xx \ps'_\xx +
\bar\ps_\xx \ps_\xx \bar\c'_\xx \c'_\xx +
\{\ps\leftrightarrow\c\}\rgt]-\\
&&-\l \sum_{\xx\in \L} \lft[\bar\c_\xx \ps_\xx \bar\c'_\xx \ps'_\xx +
\bar\c_\xx \ps_\xx \bar\ps'_\xx \c'_\xx + \{\ps\leftrightarrow\c\}\rgt] + {\rm
irr.}\;,\nn
\eea
where the {\rm irrelevant} part (irr.) is made of quartic terms with at least
one (discrete) derivative; we will discuss later on why these term are less
important. In the case of the 8V model, the second square bracket has $+\l$ in
front, rather than $-\l$.

If we set $b_\e=(q_\e+ \e q'_\e)/2$ and $d_\e=(q_\e- \e q'_\e)/2$, the
interaction with the external field is given by
\bea
&&B(\bar A) = -i \sum_{\xx\in \L\atop\e=\pm}b_\e \bar A^\e_{\xx}
\lft[\bar\ps_\xx \ps_\xx + \e\bar\ps'_\xx \ps'_\xx + \bar\c_\xx \c_\xx +
\e\bar\c'_\xx \c'_\xx
\rgt]-\nn\\
&&- i \sum_{\xx\in \L\atop\e=\pm}d_\e \bar A^\e_{\xx} \lft[ \bar\ps_\xx \ps_\xx
- \e\bar\ps'_\xx \ps'_\xx + \bar\c_\xx \c_\xx - \e\bar\c'_\xx \c'_\xx\rgt]+{\rm
irr.},\nn
\eea
where the irrelevant terms are, in this case, either quartic in the fields or
qua\-dratic  with derivatives. We remark that, if $J=J'$, then $d_\e=0$, while
$b_\e=1-\tanh(\b J) + O(\b J_4)$.

We now make another change of variables, defined by the relations
\be
\psi^\e_{\xx,+}= {\psi_\xx -\e i\psi'_\xx\over\sqrt{2}} \virg \psi^\e_{\xx,-}=
{\bar\psi_\xx -\e i \bar\psi'_\xx \over\sqrt{2}} \virg \e=\pm\;,
\ee
and the similar ones for the $\c$-variables. If we put $u=(s+s')/2$, $v=(s-s')/
2$ and $m_\e=(m_s+\e m_{s'})/2$, we get
\bea\lb{36}
&&A(\ps,m_s)+A(\ps',m_{s'}) =\\
&& = \sum_{\xx\in \L} \left\{ {u\over 4}
\lft[\ps^+_{\xx,+}\lft(\dpr^0-i\dpr^1\rgt)\ps^-_{\xx,+}+
\ps^-_{\xx,+}\lft(\dpr^0-i\dpr^1\rgt)\ps^+_{\xx,+}+{\rm c.c.}\rgt]+\right. \nn\\
&& + {u\over 4} \lft[\ps^-_{\xx,+}\lft(i\dpr^0+i\dpr^1\rgt)\ps^+_{\xx,-}+
\ps^+_{\xx,+}\lft(i\dpr^0+i\dpr^1\rgt)\ps^-_{\xx,-}+{\rm c.c.}\rgt]
+\nn\\
&&+ {v\over 4} \lft[\ps^+_{\xx,+}\lft(\dpr^0-i\dpr^1\rgt)\ps^+_{\xx,+}+
\ps^-_{\xx,+}\lft(\dpr^0-i\dpr^1\rgt)\ps^-_{\xx,+}+{\rm c.c.}\rgt] +\nn\\
&&+{v\over 4}  \lft[\ps^-_{\xx,+}\lft(i\dpr^0+i\dpr^1\rgt)\ps^-_{\xx,-}+
\ps^+_{\xx,+}\lft(i\dpr^0+i\dpr^1\rgt)\ps^+_{\xx,-}+{\rm c.c.}\rgt] -\nn\\
&&- im_+ \left. \lft[\ps_{\xx,-}^+\ps^-_{\xx,+}-
\ps^+_{\xx,+}\ps^-_{\xx,-}\rgt] + im_- \lft[\ps^-_{\xx,+}\ps^-_{\xx,-}+
\ps^+_{\xx,+}\ps^+_{\xx,-}\rgt]\right\} \;,\nn
\eea
where now the c.c. operation amounts to exchange $\psi^\e_{\xx,\o}$ with
$\psi^{-\e}_{\xx,-\o}$ and $i$ with $-i$.

The interaction with the external source is
\bea\lb{BA}
&&B(\bar A) = i \sum_{\xx\in \L} (b_+ \bar A^+_\xx + d_- \bar A^-_\xx)
[\ps^+_{\xx,+}\ps^-_{\xx,-}-\ps^+_{\xx,-}\ps^-_{\xx,+}+\\
&&+\c^+_{\xx,+}\c^-_{\xx,-}-\c^+_{\xx,-}\c^-_{\xx,+}]+i \sum_{\xx\in \L} (b_-
\bar A^-_\xx + d_+ \bar A^+_\xx)\cdot\nn\\
&&\cdot [\ps^+_{\xx,+}\ps^+_{\xx,-}+\ps^-_{\xx,+}\ps^-_{\xx,-}
+\c^+_{\xx,+}\c^+_{\xx,-}+\c^-_{\xx,+}\c^-_{\xx,-}] + {\rm irr.}\;.\nn
\eea
Finally the quartic self interaction is given by
\bea\lb{int1}
&&\VV(\psi,\c) =\l\sum_{\xx\in \L} \lft[\ps^+_{\xx,+} \ps^+_{\xx,-}
\ps^-_{\xx,+} \ps^-_{\xx,-} +\c^+_{\xx,+}\c^+_{\xx,-}\c^-_{\xx,+}
\c^-_{\xx,-}\rgt]+\nn\\
&&+v(\psi,\c) +{\rm irrel.\ terms}\;,
\eea
where $v(\psi,\c)$ is a quartic interaction depending both on $\ps$ and $\c$,
which has a different expression in the AT and 8V models, as well as the
irrelevant terms.

\subsection{Multiscale integration}\lb{s2.4}

Let $\DD$ be the set of $\kk$'s such that $k_0={2\p\over L}(n_0+{1\over 2})$
and $k_1={2\p\over L}(n_1+{1\over 2})$, for  $n_0, n_1=-{L\over
2},\ldots,{L\over 2}-1$, and $L$ and even integer. Then, the Fourier transform
for the fermions with antiperiodic boundary condition is defined by
\be
\ps^\e_{\xx,\o}\defi {1\over |\L|} \sum_{\kk\in \DD}
e^{i\e\kk\xx}\hp^\e_{\kk,\o}\;.
\ee
Therefore \pref{36} can be written as
\be\lb{2.29}
A(\ps,m_s)+A(\ps',m_{s'}) = {u\over 2 |\L|}\sum_{\kk\in\DD}\F_\kk^+ S(\kk)
\F_\kk\;,
\ee
where
\bea
\F_\kk &=& (\hp^-_{\kk,+},\hp^-_{\kk,-},\hp^+_{-\kk,+},\hp^+_{-\kk,-})\;,\nn\\
\F^+_\kk &=& (\hp^+_{\kk,+},\hp^+_{\kk,-},\hp^-_{-\kk,+},\hp^-_{-\kk,-})\;,
\eea
and , if we define
\bea
\hat D_\o(\kk) &=& -i\sin k_0+\o\sin k_1\;,\nn\\
\m(\kk) &=& (\cos k_0+\cos k_1-2) + 2{1-\sqrt2+u\over u}\;,\\
\s(\kk) &=& {v\over u}(\cos k_0+\cos k_1-2) +2{v\over u}\;,\nn
\eea
the matrix $S(\kk)$ is given by
\be\lb{sds}
S(\kk) = \pmatrix{ \hat D_-(\kk)& i\m(\kk)&{v\over u} \hat D_-(\kk) &i\s(\kk)
\cr\cr -i\m(\kk)&\hat D_+(\kk)&-i\s(\kk)&{v\over u}\hat D_+(\kk) \cr\cr {v\over
u}\hat D_-(\kk)&+i\m(\kk)&\hat D_-(\kk)&i\s(\kk) \cr\cr -i\m(\kk)&{v\over
u}\hat D_+(\kk)&-i\s(\kk)&\hat D_+(\kk) }\;.
\ee

From now until the end of the section we will only consider the case $J=J'$;
some details about the anisotropic AT model are deferred to the appendix.

Hence we have $v=0$ and $\s(\kk)\=0$, so that we get the much simpler equation
\be
A(\ps,m_s)+A(\ps',m_{s'}) = - {1\over |\L|} \sum_{\kk\in\DD} \sum_{\o,\o'}
\hp^+_{\kk,\o} \hp^-_{\kk,\o'} T_{\o,\o'}(\kk)\;,
\ee
with
\be
T(\kk)= u \pmatrix{ i\sin k_0+\sin k_1& -i\m(\kk) \cr\cr i\m(\kk)& i\sin
k_0-\sin k_1 }\;.
\ee

In the same way and with similar definitions, we get also
\be
A(\c,M_s)+A(\c',M_{s'}) = - {1\over |\L|} \sum_{\kk\in\DD} \sum_{\o,\o'}
\hc^+_{\kk,\o} \hc^-_{\kk,\o'} T^{\c}_{\o,\o'}(\kk)\;,
\ee
where $T^{\c}(\kk)$ is the matrix obtained from $T(\kk)$ by substituting
$\m(\kk)$ with
\be\lb{sigc}
\m^{\c}(\kk)= (\cos k_0+\cos k_1-2) + 2{1+\sqrt2+u\over u}\;.
\ee

Hence, we can write the functional integral \pref{25} as
\be
\bar Z(\bar A)= {1\over \NN} \int\! P(d\psi) P_\c(d\c)\; e^{\QQ(\psi,\c)+
\VV(\psi,\c)+B(\bar A)}\;,
\ee
where $\NN$ is a normalization constant and $P(d\psi)$ is the (Grassmannian)
Gaussian measure with propagator
\be
g(\xx)={1\over L^2}\sum_{\kk\in \DD} e^{-i\kk\xx} T^{-1}(\kk)\;,
\ee
$P_\c(d\c)$ is the Gaussian measure with propagator $g_\c(\xx)$, which is
obtained from $g(\xx)$ by replacing $T(\kk)$ with $T^\c(\kk)$, $\QQ(\psi,\c)$
is the sum of the quadratic terms $Q(\psi,\c)$ and $Q(\psi',\c')$, represented
in terms of the new variables; $B(A)$ and $\VV(\psi,\c)$ are defined in
\pref{BA} and \pref{int1}.

If $J>0$ and $J_4$ is any real number, $u$ is a strictly increasing function of
$\tanh(\b J)$ and has range $(0,1)$, as one can check by using the definition
of $s$, see \pref{stg}. On the other hand, $\det T(\kk)=0$ only if $\kk=0$ and
$\m(\kk)=0$; hence, $g(\xx)$ has a singularity at $u=u_c=\sqrt{2}-1$, which is
an allowed value; moreover, if $\b |J_4|\ll 1$ (as we shall suppose in the
following), $u=\tanh(\b J) + O(\b J_4)$. Since we expect that the interaction
will move this singularity, it is convenient to modify the interaction by
adding a finite {\it counterterm} $i\n{1\over L^2}\sum_{\o,\kk} \o
\hp^+_{\kk,\o}\hp^-_{\kk,-\o}$, which is compensated by replacing, in the
matrix $T(\kk)$, $\m(\kk)$ with
\be
\m_1(\kk) = (\cos k_0+\cos k_1-2) + 2 (1- {u^*\over u}) \virg u^*=
\sqrt{2}-1-\n\;.
\ee
Let us call $T_1(\kk)$ the new matrix and $P_1(d\psi)$ the corresponding
measure; we get
\be\lb{2.15}
\bar Z(\bar A)= {1\over \NN_1} \int\! P_1(d\psi) P_\c(d\c)\; e^{\QQ(\psi,\c)+
\VV^{(1)}(\psi,\c)+B(\bar A)}\;,
\ee
where
\be
\VV^{(1)}(\psi,\c) = i\n{1\over L^2}\sum_{\o,\kk}\o
\hat\psi^+_{\kk,\o}\hat\psi^-_{\kk,-\o}+ \VV(\psi,\c)\;,
\ee
and $\n$ has to be determined so that the interacting propagator has an
infrared singularity at $u=u^*$; the critical temperature is uniquely
determined by the value of $u^*$.

Let us now remark that $\det T^\c(\kk)$ is strictly positive for any $\kk$, as
one can easily see by using the fact that $u\in (0,1)$. On the other hand, it
is easy to see that
\be
\QQ(\psi,\c) = - {1\over |\L|} \sum_{\kk\in\DD} \sum_{\o,\o'}
[\hp^+_{\kk,\o}\hc^-_{\kk,\o'} + \hc^+_{\kk,\o}\hp^-_{\kk,\o'}]
Q_{\o,\o'}(\kk)\;,
\ee
where $Q(\kk)$ is a matrix which vanishes at $\kk=0$. Hence, if we define
\be
\tilde\psi^+ = \psi^+ Q T_\c^{-1} \virg \tilde\psi^- = T_\c^{-1} Q \psi^-\;,
\ee
the change of variables $\c^+\rightarrow \c^+ +\tilde\psi^+$, $\c^-\rightarrow
\c^- +\tilde\psi^-$, allows us to rewrite \pref{2.15} in the form
\be\lb{2.15a}
\bar Z(\bar A)= {1\over \NN} \int\! P_{Z_1,\m_1}(d\psi) P_\c(d\c)\;
e^{\VV^{(1)}(\psi,\c-\tilde\psi)+ \tilde B(\bar A)}\;,
\ee
where $\tilde B(\bar A)$ is the functional obtained from $B(\bar A)$ by
replacing $\c$ with $\c-\tilde\ps$ and $P_{Z_1,\m_1}(d\psi)$ is the Gaussian
measure with propagator
\be\lb{lau}
g(\xx)={1\over L^2}\sum_{\kk\in \DD} e^{-i\kk\xx} (T^{(1)})^{-1}(\kk)\;,
\ee
where $T^{(1)}(\kk)=T(\kk) - Q(\kk) T_\c^{-1} Q(\kk)$. In order to agree with
the conventions about fermion models we used in our previous papers, we make
also the trivial change of variables
\be
\hat\psi^+_{\kk,\o} \rightarrow -i\o \hat\psi^+_{\tilde\kk,\o} \virg
\hat\psi^-_{\kk,\o} \rightarrow \hat\psi^-_{\tilde\kk,\o} \virg \kk=(k_0,k_1)
\virg \tilde\kk=(k_1,k_0)\;.
\ee
Hence, by an explicit calculation of $Q(\kk)$ and using the identity $u^*/u =
1-\m_1(0)/2$, one can see that  $T^{(1)}(\kk)$ is the matrix
\be\lb{cov1}
C_1(\kk) \pmatrix{ Z_1 (-i\sin k_0 +\sin k_1)+\m_{+,+}(\kk)& -\m_1
-\m_{+,-}(\kk) \cr -\m_1 - \m_{+,-}(\kk)& Z_1 (-i\sin k_0-\sin
k_1)+\m_{-,-}(\kk) \cr}
\ee
with $C_1(\kk)=1$, $\m_1=2 u^* \m_1(0)/(2 -\m_1(0))$ and $Z_1=u^*$; moreover
$\m_{+,+}(\kk) = -\m_{-,-}(\kk)^*$ is an odd function of $\kk$ of the form
$\m_{+,+}(\kk) = 2 u^* \m_1(0) (-i\sin k_0 + \sin k_1)/(4-2\m_1(0)) +
O(|\kk|^3)$, while $\m_{+,-}(\kk)$ is a real even function, of order $|\kk|^2$,
which vanishes only at $\kk=0$. Finally, $\det T^{(1)}(\kk)\ge C(2-\cos k_0 -
\cos k_1)$, so that $P_{Z_1,\m_1}(d\psi)$ has the same type of infrared
singularity as $P_1(d\psi)$.

The fact that $\det T_\c(\kk)$ is strictly positive implies that $g_\c(\xx)$ is
an exponential decaying function; hence, we can safely perform the integration
over the field $\c$ in \pref{2.15a}. The result can be written in the following
form (see Lemma 1 of \cite{[M1]})
\be\lb{3.1}
\bar Z(\bar A)\= e^{\SS(\bar A)} =\int P_{Z_1,\m_1}(d\psi) e^{L^2\NN^{(1)}+
\bar\VV^{(1)}(\psi) + B^{(1)}(\bar A)}\;,
\ee
where $\NN^{(1)}$ is a constant and the {\it effective potential}
$\bar\VV^{(1)}(\ps)$ can be represented as
\be\lb{3.2aaa}
\bar\VV^{(1)}= \sum_{n\ge 1}\sum_{\underline\a,\underline\o,\underline\e}
\sum_{\xx_1,..,\xx_n}
W_{\underline\o,\underline\a,\underline\e,2n}(\xx_1,..,\xx_{2n})
\partial^{\a_1}\psi^{\e_1}_{\xx_1,\o_1}...
\partial^{\a_{2n}}\psi_{\xx_{2n},\o_{2n}}^{\e_{2n}}\;.
\ee
The kernels $W_{\underline\o,\underline\a,\underline\e,2n}$ in the previous
expansions are analytic functions of $\l$ and $\n$ near the origin; if we
suppose that $\n=O(\l)$, their Fourier transforms satisfy, for any $n\ge 1$,
the bounds, see \cite{[M1]}
\be
|\widehat W_{\underline\a,\underline\o,\underline\e,2n}(\kk_1,...\kk_{2n-1})|
\le L^2 C^n |\l|^{n}\;.
\ee
A similar representation can be written for the functional of the external
field $B^{(1)}(\bar A)$.

As explained in detail in \cite{[M1]}, the symmetries of the two models we are
considering
imply that, in the r.h.s. of \pref{3.2aaa}, there are no local terms
quadratic in the field, which are relevant or marginal, except those which are
already present in the free measure and are all marginal. It follows that the
integration in \pref{3.1} can be done by iteratively integrating the fields
with decreasing momentum scale and by moving to the free measure all the
marginal terms quadratic in the field. We introduce a scaling parameter $\g=2$,
a decomposition of the unity $1=f_1+ \sum_{h=-\io}^0 f_h(\kk)$, with $f_h(\kk)$
a function with support $\{\g^{h-1}\pi/4\le |\kk|\le\g^{h+1}\pi/4\}$, and the
corresponding decomposition of the field $\psi=\sum_{j=-\io}^1 \psi^{(j)}$. If
the fields $\psi^{(1)},..,\psi^{(h+1)}$ are integrated, we get
\be\lb{th1}
e^{\SS(\bar A)} = e^{S^{(h)}(\bar A)} \int P_{Z_h,\m_h}(d\psi^{(\le h)})
e^{\VV^{(h)}(\sqrt {Z_h}\psi^{(\le h)}) + \BB^{(h)}(\sqrt {Z_h}\psi^{(\le
h)},\bar A)}\;,
\ee
where $\psi^{(\le h)} = \sum_{j=-\io}^h \psi^{(j)}$ and $P_{Z_h,\m_h}(d\psi)$
is the Gaussian measure with the propagator obtained from \pref{lau} by
replacing in \pref{cov1} $C_1(\kk)$ with $C_h(\kk)=[\sum_{k=-\io}^h
f_h(\kk)]^{-1}$, $\m_1$ with $\m_h$, $Z_1$ with $Z_h$ and the functions
$\m_{\s,\s'}(\kk)$ with similar functions $\m^{(h)}_{\s,\s'}(\kk)$ (which turn
out to be negligible for $h\to -\io$, as a consequence of the following
analysis). The {\it effective interaction } $\VV^{(h)}(\psi)$ can be written as
\be\lb{5.8}
\VV^{(h)}(\psi)= \g^h \n_h F_\n^{(h)} +\l_h F_\l^{(h)}+R^{(h)}(\psi)\equiv
\LL\VV^{(h)}(\psi)+R^{(h)}(\psi) \;,
\ee
where $\n_h$ and $\l_h$ are suitable real numbers,
\bea\lb{2.112}
F_\n^{(h)} &=& {1\over L^2}\sum_\o \sum_{\kk}
\hat\psi^{(\le h)+}_{\kk,\o} \hat\psi^{(\le h)-}_{\kk,-\o}\;,\\
F_\l^{(\le h)} &=& {1\over L^8} \sum_{\kk_1,...,\kk_4} \hat\psi^{(\le
h)+}_{\kk_1,+} \hat\psi^{(\le h)+}_{\kk_3,-} \hat\psi^{(\le h)-}_{\kk_2,+}
\hat\psi^{(\le h)-}_{\kk_4,-}\d(\kk_1-\kk_2+\kk_3-\kk_4)\;, \nn\eea
and $R^{(h)}(\psi)$ is expressed by a sum over monomials similar to
\pref{3.2aaa}, with $2n+\a_1+..+\a_{2n}> 4$ ; the kernels are bounded if
$\sup_{k\ge h} (|\l_k|+|\n_k|)$ is small enough. According to power counting,
$F_\n$ is relevant, $F_\l$ is marginal while all terms in $R^h$ are irrelevant.
Moreover
\bea \lb{hhj}
&&\BB^{(h)}(\sqrt{Z_h}\psi^{(\le h)},\bar A)=\sum_{\e,\xx} Z_h^{(\e)}  \bar
A^\e_\xx O^{(\le h)\e}_\xx +R_1^{(h)}(\psi^{(\le h)},\bar A)\equiv\\
&& \LL\BB^{(h)}(\sqrt{Z_h}\psi^{(\le h)},\bar A)+R_1^{(h)}(\psi^{(\le h)}, \bar
A)\;,\nn
\eea
where
\bea\lb{curr}
O^{(\le h)+}_\xx &=& \ps^{(\le h)+}_{\xx,+} \ps^{(\le h)-}_{\xx,-} +
\ps^{(\le h)+}_{\xx,-} \ps^{(\le h)-}_{\xx,+}\;,\\
O^{(\le h)-}_\xx &=& i[ \ps^{(\le h)+}_{\xx,+} \ps^{(\le h)+}_{\xx,-} +
\ps^{(\le h)-}_{\xx,+} \ps^{(\le h)-}_{\xx,-}]\;,\nn
\eea
and $R_1^{(h)}(\psi^{(\le h)},\bar A)$ is a sum of irrelevant terms. Note that
many other possible local marginal or relevant terms could be generated in the
RG integration, which are however absent due to the symmetry of the problem, as
proved in \cite{[M1]}, App.F (see also \cite{[GM1]}, \S A2.2). The above
integration procedure is done till the scale $h^*$ defined as the maximal $j$
such that $\g^j\le |\m_j|$, and the integration of the fields $\psi^{(\le
h^*)}$ can be done in a single step. Roughly speaking, $h^*$ defines the
momentum scale of the mass.

The propagator of the field $\psi^{(\le h)}$ can be written, for $h\le 0$, as
\be\lb{ffg}
g^{(\le h)}(\xx,\yy)=g_T^{(\le h)}(\xx,\yy)+r^{(\le h)}(\xx,\yy)\;,
\ee
where
\be\lb{ombo}
g_T^{(\le h)}(\xx,\yy)={ 1\over L^2}\sum_{\kk\in \DD} e^{-i\kk(\xx-\yy)}
{1\over Z_h} T_h^{-1}(\kk)\;,
\ee
\be
T_h(\kk)= C_h(\kk) \pmatrix{ -i k_0 +k_1 & -\m_h\cr \m_h& -i k_0-k_1\cr}\;,
\ee
and, for any positive integer $M$,
\be
|r^{(\le h)}(\xx,\yy)|\le C_M{\g^{2h}\over 1 + (\g^h|\xx-\yy|^M)}\;.
\ee
The propagator $g_T^{(h)}(\xx,\yy)$ verifies a similar bound with $\g^{h}$
replacing $\g^{2h}$. A similar decomposition can be done for
$g^{(h)}(\xx,\yy)$.

The effective couplings $\l_j$ (which, by construction, are the same in the
massless $\m=0$ or in the massive $\m\not=0$ case, see \cite{[GM1]}), satisfy a
recursive equation of the form
\be\lb{bb}
\l_{j-1}=\l_j+\b_\l^{(j)}(\l_j,...,\l_0)+\bar\b_\l^{(j)}(\l_j,\n_j;...;\l_0,\n_0)
\ee
where $\b_\l^{(j)}$, $\bar\b_\l^{(j)}$ are $\m$-independent and expressed by a
{\it convergent} expansion in $\l_j,\n_j..,\l_0,\n_0$; moreover
$\bar\b_\l^{(j)}$ vanishing if at least one of the $\n_k$ is zero. From the
decomposition \pref{ffg}, the smaller bound on propagators $r$ and because of a
special feature of the propagator $g_T$, the following property, called {\it
vanishing of the Beta function}, was proved in Theorem 2 of \cite{[BM3]} for
suitable positive constants $C$ and $\th<1$:
\be\lb{beta}
|\b_\l^{(j)}(\l_j,...,\l_j)|\le C |\l_j|^2\g^{\th j}\;.
\ee
Moreover, it is possible to prove that, for a suitable choice of $\n_1=O(\l)$,
$\n_j=O(\g^{\th j}\bar\l_j)$, if $\bar\l_j=\sup_{k\ge j}|\l_k|$, and this
implies, by the {\it short memory} property ( see for instance A4.6 of
\cite{[GM1]}), $\bar\b_\l^{(j)}=O(\g^{\th j}\bar\l_j^2)$ so that the sequence
$\l_j$ converges, as $j\to -\io$, to a smooth function $\l_{-\io}(\l)= \l
+O(\l^2)$, such that
\be\lb{2.42a}
|\l_j- \l_{-\io}| \le C\l^2 \g^{\th j}\;.
\ee
Moreover
\be\lb{ffg1}
{Z_{j-1}\over Z_j}=1+\b_z^{(j)}(\l_j,...,\l_0)+
\bar\b_z^{(j)}(\l_j,\n_j;..,\l_0,\n_0)\;,
\ee
with $\bar\b_z^{(j)}$ vanishing if at least one of the $\n_k$ is zero so that,
by $\n_j=O(\g^{\th j}\bar\l_j)$ and the short memory property,
$\bar\b_z^{(j)}=O(\l_j\g^{\th j})$. Finally
\be\lb{lau11}
\b_z(\l_j,...,\l_0)= \b_z(\l_{-\io},...,\l_{-\io})+O(\l\g^{\th h})\;,
\ee
where the last identity follows from \pref{2.42a} and the {\it short memory}
property. An important point is that the function
$\b_z(\l_{-\io},...,\l_{-\io})$ is model independent. Similar equations hold
for $Z^{(\pm)}_h,\m_h$, with leading terms again model independent.

By an explicit computation and \pref{lau11} there exist $\h_+(\l_{-\io})= c_1
\l_{-\io} +O(\l_{-\io}^2)$, $\h_{-}(\l_{-\io})= -c_1 \l_{-\io}
+O(\l_{-\io}^2)$, $\h_\m(\l_{-\io})= c_1 \l_{-\io} +O(\l_{-\io}^2)$ and
$\h_z(\l_{-\io})= c_2 \l_{-\io}^2 +O(\l_{-\io}^3)$, with $c_1$ and $c_2$
strictly positive, such that, for any $j\le 0$,
\bea\lb{lau12}
&&|\log_\g( Z_{j-1}/ Z_j) - \h_z(\l_{-\io})| \le C\l^2 \g^{\th j}\;,\\
&&|\log_\g(\m_{j-1}/ \m_j) - \h_\m(\l_{-\io})| \le
C|\l| \g^{\th j}, \nn\\
&&|\log_\g( Z^{(\pm)}_{j-1}/ Z^{(\pm)}_j) - \h_\pm(\l_{-\io})| \le C\l^2
\g^{\th j}\;.\nn
\eea
The critical indices are functions of $\l_{-\io}$ only, as it is clear from
\pref{lau11}; moreover from (6.28) ad (5.4) of \cite{[M1]},
\be \lb{pppp3}
x_\pm=1-\h_\pm+\h_z \virg \h_\m=\h_+-\h_z=1-x_+\;.
\ee
When the limit $\m\to 0$ is taken (after the limit $L\to\io$, so that all the
$Z_{\g,\g'}$ have the same limit), the multiscale integration procedure implies
the power law decay of the correlations given by \pref{xpm}.

If $\m\not=0$ (that is, if the temperature is not the critical one), the
correlations decay faster than any power with rate proportional to $\m_{h^*}$,
where, if $[x]$ denotes the largest integer $\le x$, $h^*$ is given by
\be\lb{2.45c}
h^* = \left[ {\log_\g |\m| \over 1+\h_\m} \right] \;,
\ee
so that
\be
\a={1\over 2-x_+}\;.
\ee
\section{Equivalence with an effective QFT}\lb{sec3}

\subsection{The effective QFT}

We introduce a QFT model, which has a large distance behavior of the same type
as that of the formal scaling limit of the spin models with Hamiltonian (1.1).
As a general fact, the relations between the critical indices and the coupling
depend on the regularization procedure used to define the QFT model;
the kind of regularization that we are going to use allows us
to get expressions for the critical indices, simple enough to prove the
extended scaling relations.

The QFT model is defined as the limit $N\to\io$, followed by the limit
$-l\to\io$, to be called {\it the removed cutoff limit}, of a model with an
infrared $\g^l$ and an ultraviolet $\g^N$ momentum cut-off, $-l,N\ge 0$. This
model is expressed in terms of the following Grassmann integral
\bea\lb{th1111}
&&e^{\WW_{N}(A,J,\f)}=\int\! P(d\psi^{[l, N]}) \exp\left\{ \VV^{(N)}(\psi^{[l
,N]})+
\sum_\e \int\! d\xx A^{\e}_{\xx} O_{\e,\xx} +\right.\\
&&\left.+ \sum_\o \int\! d\xx\ [J_{\xx,\o} \psi^{[l,N]+}_{\xx,\o}
\ps^{[l,N]-}_{\xx,\o} + \psi^{+[l ,N]}_{\xx,\o} \f^-_{\xx,\o} + \f^+_{\xx,\o}
\psi^{[l ,N]-}_{\xx,\o}] \right\}\;,\nn
\eea
where $\xx\in\tilde\L$, a square subset of $\RRR^2$, $O^+_\xx$ and $O^-_\xx$
are defined in \pref{curr} and $P(d\psi^{[l,N]})$ is a Gaussian measure with
propagator $g_T^{[l,N]}(\xx,\yy)$ given by \pref{ombo} with $\m_h=\m,Z_h=1$ and
$C_h^{-1}(\kk)$ replaced by $C^{-1}_{l,N}(\kk)=\sum_{k=l}^N f_k(\kk)$.
The interaction is
\be\lb{gjhfk}
\VV^{(N)}(\psi)={\l_\io\over 2} \sum_{\o}\int d\xx \int d\yy v_K(\xx-\yy)
\psi^+_{\xx,\o} \psi^+_{\yy,-\o} \psi^-_{\xx,\o}\psi^-_{\yy,-\o}\;,
\ee
where $K<N$ and $v_K(\xx-\yy)$ is given by
\be
v_K(\xx-\yy)={1\over L^2}\sum_{\pp} \chi_0(\g^{-K}\pp) e^{i\pp(\xx-\yy)}\;,
\ee
$\chi_0(\pp)$ being a smooth function with support in $\{|\pp|\le 2\}$ and
equal to $1$ for $\{|\pp|\le 1\}$. The correlation functions are found by
making suitable derivatives with respect to the external fields $A_\xx$,
$J_\xx$, $\f_\xx$ and setting them equal to zero.

Note that $\lim_{K\to\io} v_K(\xx-\yy) = \d(\xx-\yy)$, so that the model
becomes the Thirring model in the limit $K\to\io$ (taken after the limit
$N\to\io$), if one also introduces an ultraviolet renormalization of the field,
$\l_\io$ and $\m$. However, in the following we shall take $K$ fixed, for
example $K=0$, so that no ultraviolet regularization is needed.

We shall study the functional $\WW_{N}(A,J,\f)$ by performing a multiscale
integration of \pref{th1111}; we have to distinguish two different regimes: the
first regime, called {\it ultraviolet}, contains the scales $h\in [K+1,N]$,
while the second one contains the scales $h\le K$, and is called {\it
infrared}.

\subsection{The ultraviolet integration}\lb{sec3.2}
We shall briefly describe how to control the integration of the ultraviolet
scales, without encountering any divergence We shall assume that the reader is
familiar with the tree expansion, as described, for example, in \cite{[BM1]},
and we only sketch the proofs, omitting many details. Moreover, for simplicity,
we shall only consider the case $A=\f=0$ and $\m=0$, but the result is valid
for the full problem; for more details in a similar case, see \cite{[M3],[M3b]}.

If the fields $\psi^{(N)},\psi^{(N-1)},...,\psi^{(h+1)}$ are integrated, we get
an expression like \pref{th1} in which the fermionic integration is
$P(d\psi^{[l, h]})$ with propagator $g^{[l,h]}_T$, and $V^{(h)}$ is sum of
integrated monomials in $m$ $\psi^+_{\xx_i,\o_i}$ variables, $i=1,\ldots,m$,
$m$ $\psi^-_{\yy_i,\o_i}$ variables and $n$ $J_{\zz_j,\o'_j}$ external fields,
$j=1,\ldots,n$, multiplied by suitable kernels $W^{(n;2m)(h)}_{\uo';\uo}
(\uz;\ux,\uy)$. These kernels are represented as power expansions in $\l$ and
$\n$, with coefficients which are finite sums of products of delta functions
(of the difference between couples of space variables) times smooth functions
of the variables which remains after the constraints implied by the the delta
functions are taken into account. With an abuse of notation, we shall denote by
$\int\! d\uz d\ux d\uy\; \lft|W^{(n;2m)(k)}_{\uo';\uo}(\uz;\ux,\uy)\rgt|$ the
expansion which is obtained by summing, for each coefficient, the $L^1$ norm of
these smooth functions. We introduce the following norm
\be\lb{norm}
\|W^{(n;2m)(k)}_{\uo';\uo}\| \defi \frac{1}{|\tilde\L|} \int\! d\uz d\ux d\uy\;
\lft|W^{(n;2m)(k)}_{\uo';\uo} (\uz;\ux,\uy)\rgt|\;.
\ee

\begin{theorem}\lb{t3.2}
If $\l_\io$ is small enough, there exist two constants $C_1>1$ and $C_2$, such
that, if $K\le h\le N$, the relevant or marginal contributions to the effective
potential satisfy the bounds:
\bea
\lb{hb1} &&\|W^{(0;2)(h)}_{\o}\| \le C_1|\l_\io|\g^{h}\g^{-2(h-K)}\;,\\
\lb{hb2} &&\|W^{(1;2)(h)}_{\o';\o}-\d_2\d_{\o,\o'}\| \le C_2|\l_\io|\g^{-(h-K)}\;,\\
\lb{hb3} &&\|W^{(0;4)(h)}_{\o,\o'}- \l_\io v\d_4 \d_{\o,-\o'}\| \le
C_2|\l_\io|^2 \g^{-(h-K)}\;,
\eea
where $\d_2(\zz;\xx,\yy) \= \d(\zz-\xx) \d_(\zz-\yy)$ and $v\d_4(\xx_1,\xx_2,
\yy_1,\yy_2) \= \d(\xx_1-\yy_1)v_K(\xx_1-\xx_2) \d(\xx_2-\yy_2)$.
\end{theorem}
{\bf\0Proof.} The proof is by induction: we assume that the bounds
\pref{hb1}-\pref{hb3} hold for $h:k+1\le h\le N$ (for $h=N$ they are true with
$C_1=C_2=0$) and we prove them for $h=k$.

The starting point is the following remark. Suppose that we build the tree
expansion, by defining the {\it localization operation} so that it acts as the
identity on the relevant or marginal terms, that is $W^{(0;2)(h)}_{\o}$,
$W^{(1;2)(h)}_{\o';\o}$ and $W^{(0;4)(h)}_{\o,\o'}$, while it annihilates, as
always, all the other contributions to the effective potential. Then, it is
easy to see that the inductive assumption implies the following ``dimensional''
bound, for $\l_\io$ small enough:
\be\lb{pc1}
\|W^{(n;2m)(k)}_{\uo';\uo}\| \le C^{n+d_{n,m}}|C_1 \l_\io|^{d_{n,m}}
\g^{k(2-n-m)}\;,
\ee
where $d_{n,m}=\max\{m-1,0\}$, if $n>0$, and $d_{n,m}=\max\{m-1,1\}$, if $n=0$,
and $C$ is a suitable constant larger, at least, of $\g$. In fact, the
localization procedure and the bounds \pref{hb1}-\pref{hb3} imply that all the
tree vertices have positive dimension and there are three types of endpoints,
associated to $W^{(0;2)(h)}_{\o}$, $W^{(1;2)(h)}_{\o';\o}$,
$W^{(0;4)(h)}_{\o,\o'}$, which contribute (up to dimensional factors and for
$\l_\io$ small enough) a factor $C_1 |\l_\io|$, $1+C_2|\l_\io|\le C$ and
$|\l_\io|[1+C_2|\l_\io|]\le C_1 |\l_\io|$, respectively. Note that the
condition $C>\g$ comes from the bound of the trivial tree (that with only one
endpoint) contributing to the tree expansion of $W^{(0;2)(k)}_{\o}$.

We need to improve the bound \pref{pc1} when $2-n-m\ge 0$. We can write, by
using the properties of the fermionic truncated expectations and the fact that,
by the oddness of the free propagator, $W^{(1;0)}_\o(\kk)=0$,
\bea\lb{111b}
&&W^{(0;2)(k)}_{\o}(\xx,\yy) =\\
&&= \l_\io \int\!d\ww d\ww'\; v_{K}(\xx-\ww)
g_\o^{[k+1,N]}(\xx-\ww')W^{(1;2)(k)}_{-\o;\o}(\ww;\ww',\yy)\;,\nn
\eea
\insertplot{190}{70}
{\ins{6pt}{36pt}{$\o$}
\ins{59pt}{36pt}{$\o$}
\ins{12pt}{28pt}{$\xx$}
\ins{53pt}{28pt}{$\yy$}

\ins{78pt}{31pt}{$=$}

\ins{98pt}{36pt}{$\o$}
\ins{178pt}{36pt}{$\o$}
\ins{102pt}{28pt}{$\xx$}
\ins{126pt}{16pt}{$\ww'$}
\ins{128pt}{57pt}{$\ww$}
\ins{172pt}{28pt}{$\yy$}

%

}%
{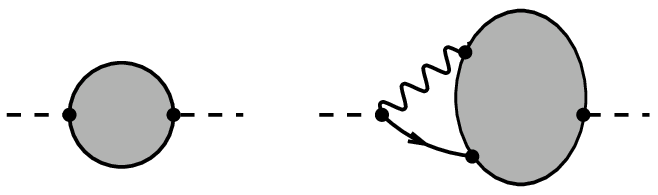}{\lb{p1}: Graphical representation of \pref{111b}}{0}
which can be bounded, by using \pref{pc1}, as
\bea\lb{111c}
&& \|W^{(0;2)(k)}_\o\| \le |\l_\io| \|v_K\|_{L^\io} \|W_{-\o;\o}^{(1;2)(k)}\|
\sum_{j=k+1}^N \|g^{(j)}_\o\|_{L^1}\le\nn\\
&& \le {c_1\over 1-\g^{-1}}\g^{2K} C |\l_\io| \g^{-k} \le C_1 |\l_\io| \g^{k}
\g^{-2(k-K)}\;,
\eea
where, for example, $C_1= \max\{2, {c_1\over 1-\g^{-1}} C\}$; hence \pref{hb1}
is proved. Note that the condition $C_1\ge 2$ is introduced only because $C_1$
is the same constant appearing in \pref{pc1}.
\insertplot{310}{140}
{\ins{6pt}{107pt}{$\o$}
\ins{12pt}{98pt}{$\xx$}
\ins{59pt}{107pt}{$\o$}
\ins{53pt}{98pt}{$\yy$}
\ins{22pt}{129pt}{$\o'$}
\ins{41pt}{120pt}{$\zz$}

\ins{85pt}{105pt}{$-\ \d_{\o',\o}$}
\ins{127pt}{107pt}{$\o$}
\ins{122pt}{95pt}{$\zz=\xx=\yy$}

\ins{170pt}{102pt}{$=$}

\ins{197pt}{107pt}{$\o$}
\ins{203pt}{97pt}{$\xx$}
\ins{230pt}{86pt}{$\uu$}
\ins{274pt}{96pt}{$\o$}
\ins{265pt}{86pt}{$\yy$}
\ins{230pt}{130pt}{$\ww$}
\ins{274pt}{133pt}{$\o'$}
\ins{265pt}{130pt}{$\zz$}
\ins{245pt}{143pt}{(a)}

\ins{25pt}{43pt}{$+$}

\ins{19pt}{17pt}{$\o$}
\ins{24pt}{7pt}{$\xx=\yy$}
\ins{39pt}{26pt}{$\ww$}
\ins{22pt}{67pt}{$\o'$}
\ins{41pt}{57pt}{$\zz$}
\ins{55pt}{60pt}{(b)}

\ins{77pt}{43pt}{$+$}

\ins{101pt}{17pt}{$\o$}
\ins{107pt}{7pt}{$\xx$}
\ins{133pt}{7pt}{$\uu$}
\ins{171pt}{17pt}{$\o$}
\ins{165pt}{7pt}{$\yy$}
\ins{119pt}{26pt}{$\ww$}
\ins{103pt}{67pt}{$\o'$}
\ins{121pt}{57pt}{$\zz$}
\ins{145pt}{60pt}{(c)}

\ins{190pt}{43pt}{$+\ \d_{\o',\o}$}

\ins{235pt}{27pt}{$\xx=\zz$}
\ins{230pt}{37pt}{$\o$}
\ins{262pt}{28pt}{$\uu$}
\ins{303pt}{37pt}{$\o$}
\ins{297pt}{28pt}{$\yy$}
\ins{255pt}{60pt}{(d)}
}%
{p2}{\lb{p2}: Graphical representation of
$W^{(1;2)(k)}_{\o';\o}(\zz;\xx,\yy)$}{0}

Let us now consider $W^{(1;2)(k)}_{\o';\o}(\zz;\xx,\yy)$ and note that it can
be decomposed as the sum of the five terms in Fig.\ref{p2},
The term denoted by $(a)$ in Fig.\ref{p2} can be bounded as
\bea
\hspace{-0.5cm} \|W^{(1;2)(k)}_{(a);\o';\o}\| \le |\l_\io| \|v_K\|_{L^\io}
\|W^{(2;2)(k)}_{\o',-\o;\o}\| \sum_{j=k+1}^N \|g^{(j)}_\o\|_{L^1} \le C C_1
|\l_\io| \g^{-2(k-K)}\;.
\eea
The bounds for the graphs $(c)$ and $(d)$ are an easy consequence of the the
bound for $W^{(0;2)(k)}_\o$.

In order to obtain an improved bound also for the graph $(b)$ of Fig. \ref{p2},
we need to further expand $W^{(2;0)(k)}_{\o,\o'}$ as done in Fig \ref{p3}, if
we suppose that the arrows in the fermion lines of graph $(b2)$ can be
reversed.
\insertplot{270}{100}
{\ins{1pt}{82pt}{$\o$}
\ins{3pt}{65pt}{$\xx$}
\ins{25pt}{65pt}{$\ww$}
\ins{50pt}{93pt}{$\uu'$}
\ins{50pt}{54pt}{$\uu$}
\ins{84pt}{85pt}{$\o'$}
\ins{78pt}{65pt}{$\zz$}
\ins{13pt}{93pt}{(b)}

\ins{116pt}{72pt}{$=$}
\ins{130pt}{74pt}{$\d_{\o',-\o}$}

\ins{155pt}{82pt}{$\o$}
\ins{157pt}{63pt}{$\xx$}
\ins{174pt}{63pt}{$\ww$}
\ins{232pt}{82pt}{$-\o$}
\ins{226pt}{63pt}{$\zz$}
\ins{170pt}{93pt}{(b1)}

\ins{-10pt}{24pt}{$+$}

\ins{1pt}{32pt}{$\o$}
\ins{3pt}{14pt}{$\xx$}
\ins{23pt}{14pt}{$\ww$}
\ins{35pt}{36pt}{$\uu'$}
\ins{61pt}{18pt}{$\zz'$}
\ins{82pt}{6pt}{$\uu$}
\ins{82pt}{46pt}{$\ww'$}
\ins{100pt}{35pt}{$\o'$}
\ins{92pt}{14pt}{$\zz$}
\ins{13pt}{49pt}{(b2)}

\ins{122pt}{24pt}{$+$}

\ins{137pt}{32pt}{$\o$}
\ins{139pt}{13pt}{$\xx$}
\ins{159pt}{13pt}{$\ww$}
\ins{207pt}{13pt}{$\uu$}
\ins{217pt}{16pt}{$\zz'$}
\ins{256pt}{32pt}{$\o'$}
\ins{250pt}{13pt}{$\zz$}
\ins{148pt}{43pt}{(b3)}
}%
{p3a}{\lb{p3}: Graphical representation of graph (b) in Fig.\ref{p2} }{0}

The bound for the graph $(b2)$ can be done by using the previous arguments. We
can write
\bea
&& W^{(1;2)(k)}_{(b2)\o';\o}(\zz;\xx,\yy) = \l_\io^2 \d(\xx-\yy)\int\! d\ww
d\uu' d\zz'\ v_K(\xx-\ww) v_K(\uu'-\zz')\cdot\nn\\
&&\cdot\; \int\! d\uu  d\ww'\ g_{\o}^{[k+1,N]}(\ww-\uu) g_{\o}^{[k+1,N]}(\uu'-\ww)
g_{\o}^{[k+1,N]}(\ww'-\uu') \cdot\nn\\
&& \cdot W^{(2;2)(k)}_{\o',\o;-\o}(\zz,\zz';\ww',\uu)\;.
\eea
In order to get the right bound, it is convenient to decompose the three
propagators $g_\o$ into scales and then bound by the $L^\io$ norm the
propagator of lowest scale, while the two others are used to control the
integration over the inner space variables through their $L^1$ norm. Hence we
get:
\bea\lb{27}
&& \|W^{(1;2)(k)}_{(b2)\o';\o}\|\le |\l_\io|^2 \|v_K\|_{L^\io} \|v_K\|_{L^1}
\|W^{(2;2)(k)}_{\o',-\o;\o}\| \cdot\\
&&\cdot 3!\sum_{k+1 \le i' \le j \le i \le N} \|g^{(j)}_{\o}\|_{L^1}
\|g^{(i)}_{\o}\|_{L^1} \|g^{(i')}_{\o}\|_{L^\io} \le C_3 |\l_\io|^2 \g^{-2(k-K)}\;.
\eea
for some constant $C_3$.

The bound of $(b1)$ and $(b3)$ requires a new argument, based on a cancelation
following from the particular form of the free propagator. Let us consider, for
instance, $(b1)$:
\bea
&& W^{(1;2)(k)}_{(b1)\o';\o}(\zz;\xx,\yy) =\nn\\
&& = \l_\io \d_{\o',-\o}\d(\xx-\yy)\int\!d\ww\ v_K(\xx-\ww)
\lft[g^{[k+1,N]}_{-\o}(\ww-\zz)\rgt]^2\;.
\eea
On the other hand, since the cutoff function $C_{k,N}(\kk)$ is symmetric in the
exchange between $k_0$ and $k_1$, it is easy to see that $g^{[k,N]}_{\o}(x_0,x_1)=
-i\o g^{[k,N]}_{\o}(x_1, -x_0)$; hence
\be\lb{mas3}
\int\!  d\uu\ \lft[g^{[k+1,N]}_{-\o}(\uu)\rgt]^2=0\;.
\ee
It follows, by using \pref{mas3} and the identity
\be \lb{idb}
v_K(\xx-\ww)=v_K(\xx-\zz)+ \sum_{j=0,1} (z_j-w_j) \int_0^1\!\!d \t\ \big(\dpr_j
v_K\big)\big(\xx-\zz+\t(\zz-\ww)\big)\;,
\ee
that we can write
\bea
&& W^{(1;2)(k)}_{(b1)\o';\o}(\zz;\xx,\yy) = \l_\io \d_{\o',-\o}\d(\xx-\yy)
\cdot\\
&&\cdot \sum_{j=0,1} \int_0^1\!\!d \t\ \int\! d\ww\ \big(\dpr_j v_K\big)
\big(\xx-\zz+\t(\zz-\ww)\big) (z_j-w_j) \lft[g^{[k+1,N]}_{-\o}(\ww-\zz)\rgt]^2
\;.\nn
\eea
Hence,
\bea
&& \|W^{(1;2)(k)}_{(b1)\o';\o}\| \le 4|\l_\io| \sum_{i=k}^N \sum_{j=k}^i
\|g^{(j)}_{-\o}\|_{L^\io} \int\!d\xx\ \big| (\dpr_j v_K)(\xx) \big|\cdot\\
&& \cdot \int\!d\ww\ |w_j||g^{(i)}_{-\o}(\ww)| \le C_4 |\l_\io| \g^{-(k-K)}\;.
\eea
By summing all the bounds, we see that there is a constant $C_2$ such that
\be\lb{ab}
\|W^{(1;2)(k)}_{\o';\o}-\d_{\o,\o'}\d_2\| \le C_2 |\l_\io| \g^{-(k-K)}\;,
\ee
which proves \pref{hb2}. The bound \pref{hb3} for $W^{(0;4)(k)}$ follows from
similar arguments.

\subsection{Equivalence of the spin and the QFT models}

As a consequence of the integration of the ultraviolet scales discussed in the
previous section, we can write the removed cutoffs limit of \pref{th1111}, with
$\f=J=0$ and with the choice $K=0$, as
\be\lb{221}
\lim_{l\to -\io}\lim_{N \to \io} \int P_{\m_0,Z_0}(d\psi^{(\le
0)})e^{\VV^{(0)}(\psi^{(\le 0)})+\BB^{(0)}(\psi^{(\le 0)}, A)}\;,
\ee
where the propagator of the integration measure in \pref{221} coincides with
$g_T^{(\le 0)}(\xx,\yy)$, defined in \pref{ombo}, $\LL\VV^{(0)}=\l_{0}
F_\l^{(0)}$ and $\LL\BB^{(0)}$ is defined as in \pref{hhj}; from the analysis
of the previous section it follows that $\l_{0}$ is a smooth function of
$\l_{\io}$, such that $\l_{0}=\l_{\io}+O(\l_\io^2)$.

The multiscale integration for the negative scales can be done exactly as
described in \S\ref{s2.4}, with the only difference that, by the oddness of the
free propagator, $\n_j=0$ and
\be
\l_{j-1}=\l_j+\hat \b^{(j)}_\l(\l_j,...\l_0)\;,
\ee
where, by \pref{ffg} and the short memory property,
\be
\hat \b^{(j)}_\l(\l_j,...\l_0)=\b^{(j)}_\l(\l_j,...\l_0)+ O(\bar\l_j^2\g^{\th
j})\;,
\ee
$\b^{(j)}_\l(\l_j,...\l_j)$ being the function appearing in the bound
\pref{beta}, so that we can prove in the usual way that
$\l_{-\io}=\l_{0}+O(\l_{0}^2)$; since $\l_{0}=\l_{\io}+O(\l_\io^2)$, we have
\be \l_{-\io}=h(\l_{\io}) = \l_{\io}+O(\l_\io^2)\;,
\ee
for some analytic function $h(\l_{\io})$, invertible for $\l_\io$ small enough.
Moreover
\be\lb{ffg2}
{Z^{\pm}_{j-1}\over Z^{\pm}_j}=1+\hat\b_\pm^{(j)}(\l_j,...,\l_0)\;,
\ee
with
\be\lb{dx1}
\hat\b_\pm^{(j)}(\l_j,...,\l_0)= \b^{(j)}_\pm(\l_j,...\l_0)+
O(\bar\l_j^2\g^{\th j})\;,
\ee
$\b^{(j)}_\pm$ being the functions appearing in the analogous equations for the
model of \S\ref{s2.4}. This implies that
\be\lb{dx2}
\h_\pm=\log_\g [1+\b^{(-\io)}_\pm(\l_{-\io},...\l_{-\io})]\;,
\ee
that is {\it the critical indices in the AT or 8V  or in the model
\pref{th1111} are the same as functions of $\l_{-\io}$}.

Of course $\l_{-\io}$ is a rather complex function of all the details
of the models. However, if we
call $\l'_j(\l)$ the effective couplings of the lattice model of the previous
sections, the invertibility of $h(\l_{\io})$ implies that we can choose
$\l_{\io}$ so that
\be\lb{dx3}
h({\l_{\io}})=\l'_{-\io}(\l)\;.
\ee
With this choice of $\l_{\io}(\l)$, the critical indices are the same, as they
depend only on $\l_{-\io}$; the rest of this chapter is devoted to the proof
that the critical indices have, as functions of $\l_{\io}$, simple expressions,
which imply the scaling relations in the main theorem.
%
%
\*
{\bf Remark} \pref{dx2} and \pref{dx3} play a central role in our analysis;
they say that the critical indices of the spin lattice models (1.1) are equal
to the ones of the QFT model \pref{th1111}, provided that its coupling is
chosen properly; such a model is defined in the continuum but the non locality
of the interaction has the effect that no ultraviolet divergences are
generated. On the other hand, the model \pref{th1111} verifies extra
symmetries, involving Ward Identities and closed equation, which allow us to
derive simple expressions for the indices in terms of $\l_\io$, as we will see
in the following sections.

\subsection{Ward Identities}\lb{sec3.4}
We consider the case $\m=0$ and we call $D_\o(\kk)=-i k_0+\o k$. We shorten the
notation of  $\WW_N(0,J,\h)$ into $\WW_N(J,\h)$. By the change of variables
$\psi^\pm_{\xx,\o}\to e^{\pm i\a_{\xx,\o}}\psi^\pm_{\xx,\o}$ we obtain the
identity
\bea\lb{grez}
&& D_\o(\pp){\partial \WW_{N}\over \partial \hJ_{\pp,\o}}(0,\h) -\n\;
\hv_K(\pp) D_{-\o}(\pp){\partial \WW_{N}\over \partial \hJ_{\pp,-\o}}(0,\h)=\\
&&= \int\!{d\kk\over (2\p)^2} \left[\hh^+_{\kk+\pp,\o} {\partial \WW_{N}\over
\partial \hh^+_{\kk,\o}}(0,\h)- {\partial \WW_{N}\over \partial
\hh^-_{\kk+\pp,\o}}(0,\h) \hh^-_{\kk,\o}\right] + {\partial \WW_\AAA\over
\partial \ha_{\pp,\o}}(0,0,\h)\;,\nn
\eea
where $\n$ is a constant to be chosen later,
\bea \lb{h11}
e^{\WW_\AAA (J,\a,\h)} =\int\! P(d\ps^{[l,N]}) e^{\VV^{(N)}(\psi^{[l ,N]})+
\sum_{\o} \int\! d\xx\ J_{\xx,\o}\psi^{[l,N]+}_{\xx,\o}\ps^{[l,N]-}_{\xx,\o}}
\cr\cr \cdot e^{\sum_\o\int d\xx
[\psi^{[l,N]+}_{\xx,\o}\h^-_{\xx,\o}+\h^+_{\xx,\o}\psi^{[l,N]-}_{\xx,\o}]}
e^{\lft[\AAA_0 - \n \AAA_{-}\rgt]\lft(\a,\psi^{[l,N]}\rgt)}\;,\nn
\eea
\bea
&& \AAA_0 (\a,\ps) \defi \sum_{\o=\pm}\int\! {d\qq\;d\pp\over (2\p)^4}\
 C_\o(\qq,\pp)\ha_{\qq-\pp,\o}\hp^+_{\qq,\o}\hp^-_{\pp,\o}\;,\\
&&\hspace{-0.5cm} \AAA_-(\a,\ps) \defi \sum_{\o=\pm}\int\! {d\qq\;d\pp\over
(2\p)^4}\ D_{-\o}(\pp-\qq)\hat v_K(\pp-\qq) \ha_{\qq-\pp,\o}\hp^+_{\qq,-\o}
\hp^-_{\pp,-\o}\;,
\eea

\be
C_\o(\qq,\pp) = [\c_{l,N}^{-1}(\pp)-1] D_\o(\pp) -[\c_{l,N}^{-1}(\qq)-1]
D_\o(\qq)\;,
\ee
and $\c_{l,N}(\kk)=\sum_{k=l}^N f_k(\kk)$.

\*

\0{\bf Remark} - As explained in \S 2.2 of \cite{[BM2]}, \pref{grez} is
obtained by introducing a cut-off function $\c_{l,N}^\e(\kk)$ never vanishing
for all values of $\kk\not=0$ and equivalent to $\c_{l,N}(\kk)$ as far as the
scaling properties of the theory are concerned; $\e$ is a small parameter and
$\lim_{\e\to 0^+} \c_{l,N}^\e(\kk)=\c_{l,N}(\kk)$. This further regularization
(to be removed before taking the removed cutoffs limit) ensures that  the
identity $[(\c_{l,N}^\e)^{-1}(\kk)-1]\c_{l,N}^\e(\kk) =1-\c_{l,N}^\e(\kk)$ is
satisfied for all $\kk\not=0$. When this further regularization is removed, all
the quantities we shall study have a well defined expression. \*
The two equations obtained from \pref{grez} by putting $\o=\pm 1$ can be solved
w.r.t. $\partial e^{\WW_{N}}/ \partial \hJ_{\pp,\o}$ and, if we define
\bea
&&a(\pp) = {1\over 1- \n\,\hv_K(\pp)} \virg \bar a (\pp)= {1\over 1+
\n\,\hv_K(\pp)}\;,\nn\\
\lb{Aeps} && A_\e(\pp) = {a(\pp) + \e\bar{a}(\pp)\over2}\;,
\eea
we obtain the identity
\bea\lb{WT1} && {\partial e^{\WW_{N}}\over \partial
\hJ_{\pp,\o}}(0,\h) -\sum_{\o'}{A_{\o\o'}(\pp)\over D_{\o}(\pp)} {\partial
e^{\WW_\AAA}\over \partial \ha_{\pp,\o'}}(0,0,\h) =\\
&&=\sum_{\o'} {A_{\o\o'}(\pp)\over D_{\o}(\pp)} \int\! {d \kk\over (2\p)^2}\
\left[\hh^+_{\kk+\pp,\o'} {\partial e^{\WW_{N}}\over
\partial \hh^+_{\kk,\o'}}(0,\h)- {\partial e^{\WW_{N}}\over
\partial \hh^-_{\kk+\pp,\o'}}(0,\h) \hh^-_{\kk,\o'}\right]\;.\nn
\eea

Given a correlation function with $m$ external fields of momenta $\kk_1,
\ldots, \kk_m$, we shall say that its {\it external momenta are non
exceptional}, if, for any subset $I$ of $\{1,\ldots,m\}$, $\sum_{i\in I}
\s_i\kk_i \not=0$, where $\s_i=+1$ for the incoming momenta and $\s_i=-1$ for
the outcoming momenta. Note that our definitions are such that $\h^+$ is an
incoming field, while $\h^-$, $J$ and $\a$ are outcoming.

An important role in this paper will have the following lemma, which was
already proved in \cite{[M3]}.
\begin{lemma}
If $\l_\io$ is small enough, there exists a choice of $\n$, independent of $l$
and $N$, such that
\be\lb{61bb}
\n ={\l_\io\over 4\pi}
\ee
and, in the limit of removed cut-offs,
\be\lb{as}
\sum_{\o'}{A_{\o\o'}(\pp)\over D_{\o}(\pp)} {\partial e^{\WW_\AAA}\over
\partial \ha_{\pp,\o'}}(0,0,\h)=0\;,
\ee
in the sense that the correlation functions generated by deriving w.r.t. $\h$
the l.h.s. of \pref{as} vanish in the limit of removed cutoffs, if the external
momenta are non exceptional.
\end{lemma}

{\bf\0Proof.} We sketch here the proof, as it will be useful in the following,
referring for more details to \cite{[M3]} (see also \cite{[FM]} and
\cite{[BFM1],[BM3]}). The starting point is the remark that $\WW_\AAA(\a,0,\h)$
is very similar to $\WW_N(J,\h)$, see \pref{th1111}, the difference being that
$\int J_{\xx,\o}\psi^+_{\xx,\o}\psi^-_{\xx,\o}$ is replaced by $\AAA_0 - \n
\AAA_-$. A crucial role in the analysis is played by the function
$C_\o(\pp,\qq)$ appearing in the definition of $\AAA_0$; this function is very
singular, but it appears in the various equations relating the correlation
functions only through the regular function
\bea\lb{mjmj}
\hU^{(i,j)}_\o(\qq+\pp,\qq) \defi \tilde\c_N(\pp) C_\o(\qq+\pp,\qq)
\hg^{(i)}_\o(\qq+\pp) \hg^{(j)}_\o(\qq)\;,
\eea
where $\tilde\c_N(\pp)$ is a smooth function, with support in the set $\{|\pp|
\le 3\g^{N+1}\}$ and equal to $1$ in the set $\{|\pp| \le 2\g^{N+1}\}$; we can
add freely this factor in the definition, since $\hU^{(i,j)}_\o(\qq+\pp,\qq)$
will only be used for values of $\pp$ such that $\tilde\c_N(\pp)=1$, thanks to
the support properties of the propagator. It is easy to see that
$\hU^{(i,j)}_\o$ vanishes if neither $j$ nor $i$ equals $N$ or $l$; this has
the effect that at least one of the fields in $\AA_0$ has to be integrated at
the $N$ or $l$ scale.

As a matter of fact, the terms in which at least one field is integrated at
scale $l$ are much easier to analyze, see below. In order to study the others,
it is convenient to introduce the function $\hS_{\bar\o,\o}^{(i,j)}$ defined by
the equation
\bea\lb{91}
\hU_\o^{(i,j)}(\qq+\pp,\qq) =\sum_{\bar\o}D_{\bar\o}(\pp)
\hS_{\bar\o,\o}^{(i,j)}(\qq+\pp,\qq)\;.
\eea
One can show that, if we define
\be
S_{\bar\o,\o}^{(i,j)}(\zz;\xx,\yy) =\int\!{d\pp\;d\qq\over (2\p)^4}\;
e^{-i\pp(\xx-\zz)}e^{i\qq(\yy-\zz)}\hS_{\bar\o,\o}^{(i,j)}(\pp,\qq)\;,
\ee
then, given any positive integer $M$, there exists a constant $C_M$ such that,
if $j>l$,
\be\lb{61}
|S_{\bar\o,\o}^{(N,j)}(\zz;\xx,\yy)| \le C_M {\g^{N}\over 1+[\g^N|\xx-\zz|]^M}
{\g^{j}\over 1+[\g^j|\yy-\zz|]^M}\;,
\ee
a bound which is used to control the renormalization of the marginal terms
containing a vertex of type $\AAA_0$.
%
%
We choose $\n$ as given by
\be\lb{61b}
\n= \l_\io \sum_{i,j=l+1}^N \int\!{d\qq\over (2\p)^2}\
\hS_{-\o,\o}^{(i,j)}(\qq,\qq)\;;
\ee
by an explicit calculation one can see that, for any $l<0$ and $N>0$, $\n$
satisfies \pref{61bb}. We remark that, to get this result, it is important to
exclude from the sum in the r.h.s. of \pref{61b} the couples $(i,j)$ with one
of the indices equal to $l$; without this restriction, $\n$ would be equal to
$0$, for any $N>0$.

The fact that the external momenta are non exceptional is important to avoid
the infrared singularities of the correlation functions. This condition on the
momenta is taken into account by using the fact that, in the tree expansion of
the correlation functions, there are important constraints on the scale indices
of the trees. This allows us to safely bound the Fourier transforms of the
correlation functions by the sum over the $L^1$ norms in the coordinate space
of the contributions associated to the different trees; see \cite{[BFM1]},
\S3.1, for an example of this strategy. Moreover, the tree structure of the
expansion allows us to express the $L^1$ norm of the correlation functions in
terms of the $L^1$ norm of the effective potential on the different scales;
hence, in the following, in order to study the effect on the Fourier transform
of the correlations of the ultraviolet region, we shall study the $L^1$ norm of
the kernels in the coordinate space.

We will proceed as in the analysis of $\WW_N(J,\h)$, integrating first the ultraviolet scales
$N, N-1, \ldots, h+1$, $h\ge K$, following a procedure very similar
to the one described in \S\ref{sec3.2},
the main difference being that there appear in the effective potential new
monomials in the external field $\a$ and in $\psi$.

We consider first the terms contributing to $\WW_\AAA(\a,0,\h)$ in which at
least one of the two fields in $\AAA_0$ or $\AAA_-$ is contracted at scale $N$.
The marginal terms such that only one of these two fields is contracted are
proportional to $W^{(0;2)(k)}$, so that one can use \pref{hb1} to bound them.
Hence, we shall consider in detail only the terms such that both fields of
$\AA_0$ or $\AA_1$ are contracted and we shall call
$\hK^{(n;2m)(k)}_{\D;\o;\uo'}$ the corresponding kernels of the monomials with
$2m$ $\ps$-fields and $n$ $\a$-fields. In the case $n=1$, we decompose them as
follows:
\bea
\hK^{(1;2m)(k)}_{\D;\o;\uo'}(\pp;\uk)=
\sum_{\s}D_{\s\o}(\pp)\hW^{(1;2m)(k)}_{\D;\s,\o;\uo'}(\pp;\uk)\;,
\eea
where $\pp$ is the momentum flowing along the external $\a$-field. As in \S
\ref{sec3.2}, we have to improve the dimensional bound of
$W^{(1;2)(k)}_{\D;\s,\o;\o'}$. We can write the following identity, which is
represented the first line of Fig.\ref{p8b} in the case $\s=-1$:
\bea\lb{71ter}
&& W^{(1;2)(k)}_{\D;\s,\o;\o'}(\zz;\xx,\yy) =\sum_{i,j=k}^N\int\! d\uu d\ww\;
S^{(i,j)}_{\s\o,\o}(\zz;\uu,\ww)W^{(0;4)(k)}_{\o,\o'}(\uu,\ww,\xx,\yy)-\nn\\
&&- \n\, \d_{-1,\s}\int\! d\ww\;
v_K(\zz-\ww)W^{(1;2)(k)}_{-\o;\o'}(\ww;\xx,\yy)\;.
\eea
\insertplot{320}{230}
{\ins{57pt}{222pt}{$\o$}
\ins{57pt}{204pt}{$\zz$}
\ins{95pt}{194pt}{$\uu$}
\ins{95pt}{232pt}{$\ww$}
\ins{131pt}{207pt}{$\o'$}
\ins{120pt}{196pt}{$\xx$}
\ins{131pt}{236pt}{$\o'$}
\ins{120pt}{231pt}{$\yy$}

\ins{165pt}{216pt}{$-\ \n_N$}

\ins{202pt}{223pt}{$-\o$}
\ins{197pt}{204pt}{$\zz$}
\ins{215pt}{204pt}{$\ww$}
\ins{259pt}{208pt}{$\o'$}
\ins{245pt}{196pt}{$\xx$}
\ins{259pt}{235pt}{$\o'$}
\ins{245pt}{231pt}{$\yy$}

\ins{-5pt}{159pt}{$=$}
\ins{22pt}{168pt}{$\o$}
\ins{22pt}{146pt}{$\zz$}
\ins{72pt}{146pt}{$\uu$}
\ins{92pt}{146pt}{$\ww$}
\ins{131pt}{154pt}{$\o'$}
\ins{120pt}{140pt}{$\xx$}
\ins{131pt}{182pt}{$\o'$}
\ins{120pt}{177pt}{$\yy$}
\ins{30pt}{185pt}{(a)}

\ins{165pt}{159pt}{$-\ \n_N$}
\ins{205pt}{168pt}{$-\o$}
\ins{187pt}{146pt}{$\zz=\uu$}
\ins{217pt}{146pt}{$\ww$}
\ins{259pt}{152pt}{$\o'$}
\ins{245pt}{140pt}{$\xx$}
\ins{259pt}{182pt}{$\o'$}
\ins{245pt}{177pt}{$\yy$}
\ins{190pt}{185pt}{(b)}

\ins{7pt}{99pt}{$+$}
\ins{32pt}{106pt}{$\o$}
\ins{32pt}{88pt}{$\zz$}
\ins{62pt}{124pt}{$\uu$}
\ins{108pt}{122pt}{$\uu'$}
\ins{108pt}{78pt}{$\ww$}
\ins{87pt}{90pt}{$\ww'$}
\ins{131pt}{120pt}{$\o'$}
\ins{120pt}{115pt}{$\yy$}
\ins{131pt}{90pt}{$\o'$}
\ins{120pt}{82pt}{$\xx$}
\ins{30pt}{125pt}{(c)}

\ins{0pt}{29pt}{$+\ \d_{\o,\o'}$} \ins{42pt}{35pt}{$\o$}
\ins{42pt}{18pt}{$\zz$} \ins{82pt}{6pt}{$\ww$} \ins{122pt}{20pt}{$\o'$}
\ins{112pt}{6pt}{$\xx$} \ins{107pt}{35pt}{$\ww'$} \ins{117pt}{56pt}{$\o'$}
\ins{88pt}{55pt}{$\uu=\yy$} \ins{30pt}{55pt}{(d)}

\ins{150pt}{29pt}{$-\ \d_{\o,\o'}$} \ins{192pt}{35pt}{$\o$}
\ins{192pt}{18pt}{$\zz$} \ins{232pt}{6pt}{$\ww$} \ins{272pt}{20pt}{$\o'$}
\ins{262pt}{6pt}{$\xx$} \ins{257pt}{35pt}{$\ww'$} \ins{247pt}{55pt}{$\uu$}
\ins{264pt}{58pt}{$\uu'$} \ins{310pt}{56pt}{$\o'$} \ins{296pt}{55pt}{$\yy$}
\ins{190pt}{55pt}{(e)}
}
{p8c}{\lb{p8b}: Graphical representation
of $W^{(1;2)(k)}_{\D;-1,\o;\o'}$}{0}

We can further decompose $W^{(1;2)(k)}_{\D;-1,\o;\o'}$ as in the last there
lines of Fig.\ref{p8b}. The term (c) can be written as
\bea
&&\l_\io \sum_{i,j=k}^N\int\! d\uu d \uu' d\ww d\ww'\;
S^{(i,j)}_{-\o,\o}(\zz;\uu,\ww)
g^{[k,N]}_\o(\uu-\uu')v_K(\uu-\ww') \cdot\nn\\
&& \cdot W^{(1;4)(k)}_{-\o;\o,\o'}(\ww';\uu',\ww,\xx,\yy)\;.
\eea
Hence, if we put $b_j(\xx)\defi \g^j/(1+[\g^j|\xx|]^3)$, we recall that
$S^{(i,j)}_{-\o,\o}$ is different from $0$ only if either $i$ or $j$ is equal
to $N$, and we use the bound \pref{61}, we see that the norm of (c) is bounded
by
\bea
&& C_3 |\l_\io| \|v_K\|_{L^\io} \sum_{i,j,m=k}^{N\, *} \int\!  d \xx d \uu'
d\ww d\ww'\; |W^{(1;4)(k)}_{-\o;\o,\o'}(\ww';\uu',\ww,\xx,\yy)| \cdot\nn\\
&&\cdot \int\!d\zz d\uu \; b_i(\zz-\ww) b_j(\zz-\uu) |g^{(m)}_\o(\uu-\uu')|\;,
\eea
where $*$ reminds that $\max\{i,j\}=N$. Since the $L^1$ and the $L^\io$ norm of
$b_j$ satisfy a bound similar to analogous bounds of $g^{(j)}_\o$, we can
proceed as in the previous section to bound $\int\!d\zz d\uu \;
 b_i(\zz-\ww) b_j(\zz-\uu)|g^{(m)}_\o(\uu-\uu')|$, by taking the $L^\io$ norm
for the factor with the smaller index and the $L^1$ norm for the other two. By
also using \pref{pc1}, we get the bound
\be
C_\th |\l_\io|^2 \g^{-2(k-K)}\g^{-\th(N-k)}\;,
\ee
for any $0<\th <1$ ($C_\th$ is divergent for $\th\to 1$). With respect to
analogous bound in \S\ref{sec3.2} ((b2) in Fig.\ref{p3}), there is an
improvement of a factor $\g^{-\th(N-k)}$.
The term (d) can be bounded by
$$C |\l_\io| \|v_K\|_{L^\io}\ \sum_{i,j= k}^{N\ *}\|b_i\|_{L^1}\
\|b_j\|_{L^1} \le C |\l_\io| \g^{-(k-K)} \g^{-(N-k)}\,;$$
for the term  (e) we get the bound $C |\l_\io|^2 \g^{-3(k-K)} \g^{-(N-k)}$. By
putting together all the previous bounds, we get
\be\lb{222}
\| (c) +(d) +(e)\| \le C_\th |\l_\io| \g^{-(k-K)} \g^{-\th(N-k)}\;.
\ee

We consider now the terms (a) and (b), whose sum can be written as
\bea\lb{75} \int\! d\uu\; \lft[\l_\io \sum_{i,j=k}^N
S^{(i,j)}_{-\o,\o}(\zz;\uu,\uu) -\n \d(\zz-\uu)\rgt] \cdot\cr\cr\cdot
\int\!d\ww\; v_K(\uu-\ww)W^{(1;2)(k)}_{-\o;\o'}(\ww;\xx,\yy)\;.
\eea
By using the identity \pref{idb}, \pref{75} can be written also as
\bea\lb{125b}
&& \left[\l_\io \sum_{i,j=k}^N\int\! d\uu \; S^{(i,j)}_{-\o,\o}(\zz;\uu,\uu)
-\n\right] \int\! d\ww\;  v_K(\zz-\ww) W^{(1;2),(k)}_{-\o;\o'}(\ww;\xx,\yy)
+\nn\\
&& + \l_\io \sum_{p=0,1}\sum_{i,j=k}^N\int\! d\uu \;
S^{(i,j)}_{-\o,\o}(\zz;\uu,\uu)(u_p-z_p) \cdot\\
&&\cdot \int_0^1\!d\t\; \int\! d\ww\; (\dpr_pv_K)(\zz-\ww+\t(\uu-\zz))
W^{(1;2),(k)}_{-\o;\o'}(\ww;\xx,\yy)\;.\nn
\eea
The latter term is again irrelevant and vanishing for $N-k\to+\io$; in fact,
its norm can be bounded by
\bea
&& 2|\l_\io| \|W^{(1;2),(k)}_{-\o;\o'}\| \, \|\dpr v_K\|_{L^1} \,
\sum_{i,j=k}^{N\
*}\int\! d\zz \; b_i(\zz-\uu) b_j(\zz-\uu)|\uu-\zz_p| \le\nn\\
&&\hspace{2cm} \le C|\l_\io| \g^{-(k-K)}\g^{-(N-k)} \;.
\eea
Contrary to what happened for the graph (b1) of Fig\ref{p3}, the contribution
of the graph (a) to the first term in the r.h. side of \pref{125b} is not zero
(that is, {\it the fermionic bubble is not vanishing}); however, in this case
its value is compensated by the graph (b), thanks to the explicit choice we
made for $\n$. Indeed we have
\be\lb{78bis}
\l_\io \sum_{i,j=k}^N\int\! d\uu \; S^{(i,j)}_{-\o,\o}(\zz;\uu,\uu)-\n= -2
\l_\io \sum_{j=l+1}^{k-1}\int\! d\uu \; S^{(N,j)}_{-\o,\o}(\zz;\uu,\uu)\;,
\ee
that easily implies that the first term in the r.h. side of \pref{125b} is
bounded by $C|\l_\io|\g^{-(N-k)}$.

Let us finally consider $W^{(1;2)(k)}_{\D;+1,\o;\o'}$, for which we can use a
graph expansion similar to that of Fig.\ref{p8b}, the only differences being
that $\n$ is replaced by $0$ and the indices $-\o$ are replaced by $\o$. Hence
a bound can be obtained with the same arguments used above, with only one
important difference: the contribution that in the previous analysis was
compensated by the graph (b) now is zero by symmetry reasons. Indeed, if we
call $\kk^*$ the vector $\kk$ rotated by $\p/2$, it is easy to see that
$\hS^{(i,j)}_{\bar\o,\o}(\kk^*,\pp^*)=-\o\bar\o\hS^{(i,j)}_{\bar\o,\o}(\kk,\pp)$,
which implies that
\be\lb{79}
\sum_{i,j=k}^N\int\! d\uu \; S^{(i,j)}_{\o,\o}(\zz;\uu,\uu)=
\sum_{i,j=k}^N\int\! {d\kk\over (2\p)^2} \; \hS^{(i,j)}_{\o,\o}(\kk,-\kk)=0\;.
\ee
We have then proved that
\be\lb{de} \|W^{(1;2)(k)}_{\D;\s,\o;\uo'}\| \le C|\l_\io|\g^{-\th(N-k)}\;,
\ee
which implies, by dimensional bounds and the short memory property, that, for
$K\le k\le N$,
\be\lb{de1}
\|W^{(1;2m)(k)}_{\D;\s,\o;\uo'}\| \le (C|\l_\io|)^m \g^{(1-m)k}\g^{-\th
(N-k)}\;.
\ee
It remains to analyze the terms contributing to $\WW_\AAA(\a,0,\h)$ in which
no one of the two fields in $\AAA_0$ is contracted at scale $N$.
If $i\ge l$ we can use the bound
\be\lb{61a}
\left| {\hU_{\o'}^{(i,l)}(\qq+\pp,\qq)\over D_\o(\pp)}\right| \le C\g^{-(i-l)}
{\g^{-l-i}\over Z_{i-1}} \virg \hbox{if\ } |\pp|\ge 2\g^{l+1} \;,
\ee
and the factor $\g^{-(i-l)}$ in the r.h.s. of this bound is an improvement
w.r.t. the dimensional bound and makes indeed irrelevant the marginal terms
containing a vertex of type $\AAA_0$, if one of the $\psi$ fields is contracted
on scale $l$ and $\pp$ has a fixed value different from $0$, as we are
supposing.

The contributions to the correlation functions generated by the l.h.s. of
\pref{as}, such that one of the $\psi$-fields in $\AAA_0$ is contracted at
scale $l$ (hence it is an external field at scale $k$), vanish in the limit
$l\to-\io$, if the momentum $\pp$ of the $\a$ field is fixed at a value
different from $0$, as we are supposing. This follows from the bound
\pref{61a}, since the value of $i$ is essentially fixed at a value of order
$\log_\g |\pp|$ and the extra factor $\g^{-(i-l)}$ vanishes for $l\to-\io$. The
correlations generated by the terms containing $W^{(1;2m)(k)}_\D$ are vanishing
in the limit of removed cut-offs, thanks to the extra factor $\g^{-\th (N-k)}$
in \pref{de1}, with respect to the dimensional one, and the short memory
property.

\subsection{Closed equations}
The Schwinger-Dyson equations for $\m=0$ are generated by the identity, see
\cite{[BFM2]},
\bea\lb{SDE}
&& D_\o(\kk) {\partial e^{\WW_{N}} \over\partial \hh^+_{\kk,\o}}(0,\h) =
\c_{l,N}(\kk) \Bigg[ \hh^-_{\kk,\o} e^{\WW_{N}(0,\h)} - \nn\\
&&- \l_\io\int\!{d\pp\over(2\p)^2}\ \hv_K(\pp){\partial^2 e^{\WW_{N}}\over
\partial\hJ_{\pp,-\o} \partial\hh^+_{\kk+\pp,\o}}(0,\h) \Bigg]\;.
\eea

By using \pref{WT1} we easily get:
\bea\lb{ce}
&& D_\o(\kk) {\partial e^{\WW_{N}} \over\partial \hh^+_{\kk,\o}}(0,\h) =
\c_{l,N}(\kk) \Bigg\{ \hh^-_{\kk,\o} e^{\WW_{N}(0,\h)} - \nn\\
&& -\l_\io\sum_{\o'} \int\! {d \pp\over (2\p)^2}\, v_K(\pp)
{A_{-\o\o'}(\pp)\over D_{-\o}(\pp)}\cdot \\
&&\cdot\int\! {d \qq\over (2\p)^2}\ \left[\hh^+_{\qq+\pp,\o'} {\partial
e^{\WW_{N}}\over \partial \hh^+_{\qq,\o'}
\partial\hh^+_{\kk+\pp,\o}}(0,\h) - {\partial e^{\WW_{N}}\over
\partial\hh^+_{\kk+\pp,\o} \partial \hh^-_{\qq+\pp,\o'}}(0,\h)
\hh^-_{\qq,\o'}\right] -\nn\\
&&-\l_\io\sum_{\o'}\int\! {d \pp\over (2\p)^2}\ v_K(\pp) {A_{-\o\o'}(\pp)\over
D_{-\o}(\pp)} {\partial^2 e^{\WW_\AAA}\over \partial
\ha_{\pp,\o'}\partial\hh^+_{\kk+\pp,\o}}(0,0,\h) \Bigg\} \;.\nn
\eea
We now want to prove that the last term in the r.h.s. of \pref{ce} is
negligible in the limit of removed cutoffs, if $\kk$ is fixed at a value far
from the cutoffs.

\begin{theorem}
In the limit of removed cutoffs, the correlation functions generated by
deriving w.r.t. $\h$ the functional
\be\lb{ff}
\sum_{\o'}\int\! {d \pp\over (2\p)^2}\ v_K(\pp) {A_{-\o\o'}(\pp)\over
D_{-\o}(\pp)} {\partial^2 e^{\WW_\AAA(0,0,\h)}\over \partial
\ha_{\pp,\o'}\partial\hh^+_{\kk+\pp,\o}}
\ee
vanish, if the external momenta are non exceptional.
\end{theorem}

{\bf Proof.} It is convenient to write \pref{ff} as $\sum_{\e=\pm} {\dpr
\WW_{T,\e}\over \dpr \hb_{\kk,\o}}(0,\h)$, where
\bea
e^{\WW_{T,\e} (\b,\h)} &=& \int\! P(d\ps^{[l,N]}) e^{\VV^{(N)}(\ps^{[l,N]}) +
\sum_\o\int d\xx [\psi^{[l,N]+}_{\xx,\o} \h^-_{\xx,\o} + \h^+_{\xx,\o}
\psi^{[l,N]-}_{\xx,\o}]} \cdot\nn\\
&\cdot& e^{\left[T^{(\e)}_{1} -\n T^{(\e)}_{-}\right]\left(\ps^{{l,N }},
\b\right)}
\eea
and
\bea\lb{80}
T^{(\e)}_{1}(\psi,\b) &=& \sum_{\o} \int\! {d\kk\;d\pp\;d\qq \over (2\p)^4}\;
\hv_K^{(\e)}(\pp) {C_{-\e\o}(\qq+\pp,\qq)\over D_{-\o}(\pp)} \cdot\nn\\
&\cdot& \hb_{\kk,\o} \hp^-_{\kk+\pp,\o} \hp^+_{\qq+\pp,-\e\o}
\hp^-_{\qq,-\e\o}\;,
\eea
\be\lb{80a}
T^{(\e)}_{-}(\psi,\b)= \sum_{\o} \int\!{d\kk\;d\pp\;d\qq\over (2\p)^4}\;
\hu_K^{(\e)}(\pp) \hb_{\kk,\o} \hp^-_{\kk+\pp,\o} \hp^+_{\qq+\pp,\e\o}
\hp^-_{\qq,\e\o}\;,
\ee
where
\be
\hv^{(\e)}_{K}(\pp)\defi v_K(\pp) \hA_{\e}(\pp) \virg \hu_K^{(\e)}(\pp) =
\hv_K^{(\e)}(\pp) \hv_K(\pp) {D_{\e\o}(\pp)\over D_{-\o}(\pp)}\;.
\ee
Note that $v^{(\pm)}_{K}(\xx)$ and $u^{(-)}_{K}(\xx)$ are smooth functions of
fast decay, hence they are equivalent to $v_K(\xx)$ in the bounds. This is not
true for $u^{(+)}_{K}(\xx)$, whose Fourier transform is bounded but
discontinuous in $\pp=0$. However, in the following we shall only need to know
that $\|u^{(+)}_{K}\|_{L^\io} \le C\g^{2K}$ and that $|\hu_K^{(+)}(\pp)| \le
|\hv_K^{(+)}(\pp) \hv_K(\pp)|$, which are easy to prove.

As in \S\ref{sec3.4}, we now perform a multiscale integration for the
ultraviolet scales $N, N-1, \ldots, k+1$, $k\ge K$, very similar to the one
described in \S\ref{sec3.2}, the main difference being that that there appear
in the effective potential new monomials in the external field $\b$ and in
$\psi$. As explained in the previous section, in order to control the Fourier
transform at non exceptional momenta of the correlation functions, it is in
general sufficient to control, in the ultraviolet region, the $L^1$ norm in
coordinate space of the kernels appearing in the effective potential. This is
in general true also in the proof of this theorem, except for a bound, where
one has to be more careful, see below.

The contributions to the correlation functions such that one of the
$\psi$-fields in $T^{(\e)}_{1}(\psi,\b)$ with momentum $\qq+\pp$ or $\qq$, see
\pref{80}, is contracted at scale $l$ (hence it is an external field at scale
$k$), vanish in the limit $l\to-\io$, if the momentum $\kk$ of $\b$ is fixed at
a value different from $0$, as we are supposing. In fact, in this case either
$|\pp|$ or $|\kk+\pp|$ is greater than $|\kk|/2$; hence, by using \pref{61a} or
the short memory property, these contributions satisfy a bound containing the
extra factor $\g^l |\kk/2|$, which vanishes for $l\to-\io$. We consider then
just the terms contributing to $\WW_{T,\e}(\b,\h)$, in which at least one of
the two $\psi$-fields in $T^{(\e)}_{1}(\psi,\b)$ with momentum $\qq+\pp$ or
$\qq$ is contracted at scale $N$. We shall call $W^{(1;2m-1)}_{T,\e;\o;\uo'}$
the corresponding kernels of the monomials with $2m-1$ $\ps$-fields and $1$
$\a$-field. We claim that
\be\lb{ti}
\|W^{(1;2m-1)(k)}_{T,\e;\o;\uo'}\| \le C \g^{(2-m)k}\g^{-\th (N-k)}\;.
\ee
By the usual arguments, this is a consequence of the improved bounds:
\bea
\lb{t1} \|W^{(1;1)(k)}_{T,\e;\o,\o}\| &\le& C |\l_\io| \g^k \g^{-\th (N-k)}
\g^{-2 (k-K)}\;,\\
\lb{t2} \|W^{(1;3)(k)}_{T,\e;\o,\uo'}\| &\le& C |\l_\io| \g^{-\th (N-k)} \;.
\eea

We prove first the bound \pref{t1}. We can write
\be
W^{(1;1)(k)}_{T,\e;\o,\o} = W^{(1;1)(k)}_{(a)T,\e;\o,\o} +
W^{(1;1)(k)}_{(b)T,\e;\o,\o}
\ee
where

\0 a) $W^{(1;1)(k)}_{(a)T,\e;\o,\o}$ is the sum over the terms such that the
field $\b$ belongs only to a $T^{(\e)}_1$-vertex, whose $\psi$-field
$\hp^+_{\qq+\pp,-\e\o}$ either is contracted with $\hp^-_{\kk+\pp,\o}$ (this
can happen only for $\e=-1$) or is connected to it through a kernel
$\hW_\o^{(0;2)(k)}(\qq+\pp)$.

\0 b) $W^{(1;1)(k)}_{(b)T,\e;\o,\o}$ is the sum over the remaining terms.

Let us consider the first term. Given $\kk$, for $N$ large enough,
$\c_{l,N}^{-1}(\kk)-1=0$; hence we can write:
\bea\lb{95}
&& \hW^{(1;1)(k)}_{(a)T,\e;\o,\o}(\kk) = \d_{\e,-1} \int\!{d\pp\over (2\p)^2}\;
{\hv^{(-1)}_K(\pp)\over D_{-\o}(\pp)} [\c_{-\io,N}(\pp+\kk)-1]\cdot\\
&& \cdot \lft[1+\hg^{[k+1,N]}_\o(\pp+\kk) \hW_\o^{(0;2)(k)}(\pp+\kk)\rgt]
\lft[1+\hg^{[k+1,N]}_\o(\kk) \hW_\o^{(0;2)(k)}(\kk)\rgt]\;.\nn
\eea
Moreover, since $\hv^{(-1)}_K(\pp)=0$ for $|\pp|\ge 2\g^K$, then
$\c_{-\io,N}(\pp+\kk)-1=0$, if $\hv^{(-1)}_K(\pp) \not=0$ and $N$ is large
enough. It follows that, given a fixed $\kk$, for $N$ large enough,
\be \hW^{(1;1)(k)}_{(a)T,\e;\o,\o}(\kk)=0\;.\ee

Let us now consider $W^{(1;1)(k)}_{(b)T,\e;\o,\o}(\xx-\yy)$, which can be
decomposed as in Fig. \ref{p9b}.
\insertplot{300}{60} {\ins{25pt}{36pt}{$\o$}
\ins{33pt}{28pt}{$\xx$}
\ins{132pt}{36pt}{$\o$}
\ins{123pt}{28pt}{$\yy$}
\ins{46pt}{56pt}{$\zz$}
\ins{80pt}{8pt}{$\ww$}

\ins{155pt}{36pt}{$-\ \n$}

\ins{184pt}{36pt}{$\o$}
\ins{193pt}{28pt}{$\xx$}
\ins{225pt}{53pt}{$\zz$}
\ins{274pt}{36pt}{$\o$}
\ins{263pt}{28pt}{$\yy$}
\ins{263pt}{28pt}{$\yy$}
\ins{225pt}{8pt}{$\ww$}
} {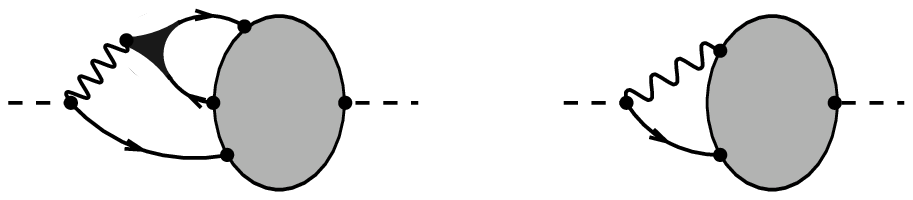}{\lb{p9b}: Graphical representation
of $W^{(1;1)(k)}_{(b)T,\e;\o,\o}$}{0}

\0 By using \pref{71ter}, it can be written as
\be
\sum_{\s} \int\! d\zz\ u^{(\e)}_K(\xx-\zz) g^{[k,N]}_\o(\xx-\ww)
W^{(1;2)(k)}_{\D;\s,-\e\o;\o}(\zz;\yy,\ww)\;,
\ee
hence its norm, by using \pref{de}, can be bounded by
\be
\|u_K^{(\e)}\|_{L^\io}\, \sum_{j=k}^N |g_\o^{(j)}|_{L^1}
\|W^{(1;2)(k)}_{\D;\s,-\e\o;\o}\| \le C |\l_\io| \g^k \g^{-2(k-K)}
\g^{-\th(N-k)}\;.
\ee

In order to prove the bound \pref{t2}, we write
\be\lb{fffg}
W^{(1;3)(k)}_{T,\e;\o;\uo'} = W^{(1;3)(k)}_{(a)T,\e;\o;\uo'}+
W^{(1;3)(k)}_{(b)T,\e;\o;\uo'}\;,
\ee
where $W^{(1;3)(k)}_{(a)T,\e;\o;\uo'}$ contains the terms in which the field
$\hp_{\kk+\pp,\o}$ of $T_1$ and $T_-$ is not contracted or is linked to a
kernel $\hW^{(0;2)(k)}_\o$, while the other terms are collected in
$W^{(1;3)(k)}_{(b)T,\e;\o;\uo'}$. Let us consider first
$W^{(1;3)(k)}_{(a)T,\e;\o;\uo'}$; its Fourier transform, if we call $\kk^+$ and
$\kk^-$ the momenta of the two fields connected to the line $u^{(\e)}_K$, can
be written as (note that $\uo'$ is of the form $(\o,\o',\o')$):
\bea
&&\hW^{(1;3)(k)}_{(a)T,\e;\o;\uo'}(\kk;\kk^+,\kk^-)= \lft[1 +
\hg^{[k+1,N]}_{\o}(\kk+\kk^+-\kk^-) \hW^{(0;2)(k)}_{\o}(\kk+\kk^+-\kk^-) \rgt]\cdot\nn\\
&&\cdot \hu^{(\e)}_K(\kk^+-\kk^-) \sum_\s
\hW^{(1;2)(k)}_{\D;\s,-\e\o,\o'}(\kk^- +\kk^+-\kk^-,\kk^-)\;.
\eea
Then, if $\e=-1$, since $\|v^{(-1)}_K\|_{L^1} \le C$, by using the bounds
\pref{de} and \pref{hb1}, we find
\be\lb{w13}
\|W^{(1;3)(k)}_{(a)T,-1;\o;\uo'}\| \le C|\l_\io| \g^{-\th (N-k) }
\ee

This bound is not true in the case $\e=+1$, where it is necessary to take
carefully into account that we are indeed bounding the Fourier transform of the
correlation functions generated by \pref{ff}, at fixed (non exceptional)
external momenta.

The terms contributing to these correlations and containing
$W^{(1;3)(k)}_{(a)T,+1;\o;\uo'}$ as a cluster can be of two different types.
There are terms such that the line corresponding to $\hp_{\kk+\pp,\o}$ is
connected to the rest of the graph only through the vertex of the field $\b$.
In this case, we have to bound an expression of the type
\be
\hu^{(+1)}_K(\kk-\qq) \hat G_1(\uk') \hat G_2(\uk'')\;,
\ee
where $\uk'$ and $\uk''$ are a set of independent external momenta,
$\qq=-\sum_i \s_i\kk'_i$, $\qq-\kk=\sum_i \s_i\kk''_i$ and $\hat G_2(\uk'')$
contains the cluster associate to $\sum_\s \hW^{(1;2)(k)}_{\D;\s,-\o,\o'}$;
this expression is bounded by $C \|G_1\|\,|\G_2\|$, the same result that we
should get in the case $\e=-1$, by bounding the full expression with the
$\|\cdot\|$ norm. Hence, the final bound is the same we would obtain by using
\pref{w13} for $\e=+1$.

We still have to consider the terms such that the line corresponding to
$\hp_{\kk+\pp,\o}$ is connected to the rest of the graph even if we erase the
vertex of the field $\b$. Now we have to bound an expression of the type
\be
\int\!{d\pp\over (2\p)^2}\; \hu^{(+1)}_K(\pp) \hg^{(j)}(\pp+\kk) \hat
G(\pp,\uk')\;,
\ee
where $\sum_i \s_i\kk_i=\kk$ and $\hat G(\pp,\uk')$ contains the cluster
associate to $\sum_\s \hW^{(1;2)(k)}_{\D;\s,-\o,\o'}$; this expression can be
bounded by $C\|\hg\|_{L^1} \|G\|$, the same result that we should get in the
case $\e=-1$, by bounding the full expression with the $\|\cdot\|$ norm.

Let us finally consider $W^{(1;3)(k)}_{(b)T,\e;\o;\uo'}$, which can be
represented as in Fig.\ref{p12}.
\insertplot{290}{120} {\ins{20pt}{120pt}{$(b1)$}
\ins{22pt}{98pt}{$\o$}
\ins{32pt}{88pt}{$\xx$}
\ins{52pt}{118pt}{$\zz$}
\ins{80pt}{70pt}{$\ww$}
\ins{129pt}{82pt}{$\o'$}
\ins{115pt}{72pt}{$\vv$}
\ins{129pt}{100pt}{$\o'$}
\ins{121pt}{86pt}{$\uu$}
\ins{129pt}{118pt}{$\o$}
\ins{121pt}{102pt}{$\yy$}

\ins{180pt}{120pt}{$(b2)$}
\ins{155pt}{96pt}{$-\ \n_N$}
\ins{182pt}{98pt}{$\o$}
\ins{192pt}{88pt}{$\xx$}
\ins{223pt}{117pt}{$\zz$}
\ins{220pt}{70pt}{$\ww$}
\ins{271pt}{82pt}{$\o'$}
\ins{257pt}{72pt}{$\vv$}
\ins{271pt}{100pt}{$\o'$}
\ins{263pt}{86pt}{$\uu$}
\ins{271pt}{118pt}{$\o$}
\ins{263pt}{102pt}{$\yy$}

\ins{6pt}{25pt}{$+$}

\ins{20pt}{50pt}{$(b3)$}
\ins{22pt}{28pt}{$\o$}
\ins{32pt}{18pt}{$\xx$}
\ins{52pt}{48pt}{$\zz$}
\ins{82pt}{52pt}{$\e\o$}
\ins{72pt}{1pt}{$\ww$}
\ins{113pt}{16pt}{$\o'$}
\ins{101pt}{3pt}{$\vv$}
\ins{113pt}{29pt}{$\o'$}
\ins{103pt}{18pt}{$\uu$}

\ins{154pt}{25pt}{$+$}

\ins{180pt}{50pt}{$(b4)$}
\ins{182pt}{28pt}{$\o$}
\ins{192pt}{18pt}{$\xx$}
\ins{212pt}{48pt}{$\zz$}
\ins{242pt}{52pt}{$\e\o$}
\ins{232pt}{1pt}{$\ww$}
\ins{273pt}{16pt}{$\o'$}
\ins{261pt}{3pt}{$\vv$}
\ins{273pt}{29pt}{$\o'$}
\ins{263pt}{18pt}{$\uu$}
} {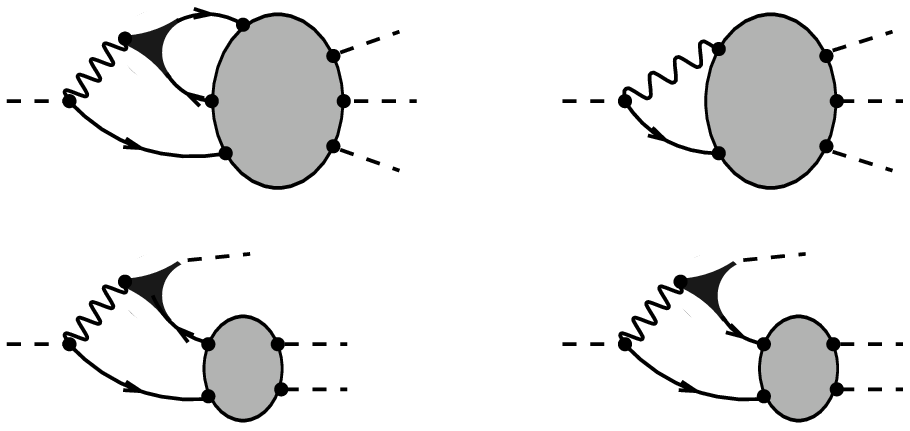}{\lb{p12}: Graphical representation
of  $W^{(1;3)(k)}_{(b)T,\e;\o;\o'}$}{0}

We can write
\bea
&& W^{(1;3)(k)}_{(b)T,\e;\o,\uo'}(\xx,\yy, \uu,\vv) = \\
&& =\int d\zz d\ww\; u_K^{(\e)}(\xx-\zz) g^{[k,N]}_\o(\xx-\ww)
W^{(1;4)(k)}_{\D,\e;\o,\o'}(\zz;\ww,\yy, \uu,\vv)\;,\nn
\eea
so that, by the bounds \pref{de}, $\|W^{(1;4)(k)}_{\D,\e;\o,\o'}\| \le C
|\l_\io| \g^{-k} \g^{-\th(N-k)}$ and $\|u^{(\e)}_K\|_{L^\io}\le C \g^{2K}$, we
get:
\be
\|W^{(1;3)(k)}_{(b)T,\e;\o;\o'}\| \le C |\l_\io|\g^{-2(k-K)}\g^{-\th(N-k)}\;.
\ee
Again, with respect to the analogous bound in \S 3.2,we have an extra factor
$\g^{-\th(N-k)}$ and this implies, proceeding for instance as in \S4.1 of
\cite{[BFM1]}, the proof of the Theorem.

\subsection{Solution of the closed equations and proof of $x_+ x_-=1$}

We want to solve the closed equations for the correlation functions
\bea
\la \psi^-_{\xx,\o}\psi^+_{\yy,\o}\ra &\defi& S_\o(\xx-\yy)\;,\\
\lb{Gom}\la \psi^-_{\xx,\o} \psi^-_{\yy,-\o} \psi^+_{\uu,-\o}
\psi^+_{\vv,\o}\ra &\defi& G_\o(\xx,\yy,\uu,\vv)\;,
\eea
in the limit of removed cutoffs. By taking in \pref{ce} one derivative w.r.t.
$\hh^-_{k,\o}$ and then putting $\h\=0$, we find
\be\lb{eqk}
 D_\o(\kk)\hS_\o(\kk)
=1 +\l_\io \int\!{d\pp\over (2\p)^2}\ \hF_{K,-}(\pp) \hS_{\o}(\kk+\pp)\;,
\ee
where
\be
\hF_{K,\e}(\pp)\defi {v_K(\pp) A_\e(\pp) \over D_{-\o}(\pp)}\;.
\ee
In the space coordinates, equation \pref{eqk} becomes
\be
\lft(\dpr_\o S_\o\rgt)(\xx) -\l_\io F_{K,-}(\xx) S_{\o}(\xx) =\d(\xx)\;,
\ee
where $\dpr_\o=\dpr_{x_0}+i\o\dpr_{x_1}$ and $F_{K,-}(\xx) = \int d\pp/(2\p)^2
e^{-i\pp\xx} \hF_{K,-}(-\pp)$. Hence, if we define
\be
\D_{\e}(\xx|\zz) = \int\!{d\kk\over (2\p)^2} \ {e^{-i\kk\xx} - e^{-i\kk\zz}
\over D_\o(\kk)} \hF_{K,\e}(-\kk)\;,
\ee
its solution is:
\be\lb{Som}
S_\o(\xx)= e^{\l_\io\D_{-}(\xx|0)}g_\o(\xx)\;.
\ee

Note that, for large  $|\xx|$, thanks to \pref{Aeps},
\be\lb{delta}
\D_{\e}(\xx|0)\sim -{A_\e(0)\over 2\p}\ln|\xx|
= -{a(0) +\e\bar a(0)\over 4\p}\ln|\xx|\;,
\ee
which implies, in particular, that the critical index $\h_z$, defined in
\pref{lau12} is equal to $[a(0) - \bar a(0)]/(4\p)$.

Let us now consider the 4-point correlation \pref{Gom}. If we take in \pref{ce}
three derivatives w.r.t. $\hh^+_{\qq,-\o}$, $\hh^-_{\kk+\qq-\bfs,\o}$ and
$\hh^-_{\bfs,-\o}$, we find:
\bea
&&D_\o(\kk)\hG_\o(\kk,\qq,\bfs) = \d(\qq-\bfs)\hS_{-\o}(\qq) +\l_\io \int
{d\pp\over (2\p)^2}\ \hF_{K,-}(\pp) \hG_\o(\kk+\pp,\qq,\bfs)+\nn\\
&&+\l_\io \int {d\pp\over (2\p)^2}\ \hF_{K,+}(\pp)
\Big[\hG_\o(\kk+\pp,\qq-\pp,\bfs) - \hG_\o(\kk+\pp,\qq,\bfs+\pp)\Big]\;,
\eea
which, in the space coordinates, becomes:
\bea
&&\lft(\dpr_\o^\xx G_\o\rgt)(\xx,\yy,\uu,\vv) = \d(\xx-\vv) S_{-\o}(\yy-\uu)+
\cr\cr &&+\l_\io \Big[F_{K,+}(\xx-\yy)- F_{K,+}(\xx-\uu) +F_{K,-}(\xx-\vv)\Big]
G_\o(\xx,\yy,\uu,\vv)\;.
\eea
By using \pref{Som}, we find that the solution of this equation is given by
\bea
G_\o(\xx,\yy,\uu,\vv) &=& e^{-\l_\io \Big[ \D_+(\xx-\yy|\vv-\yy) -
\D_+(\xx-\uu,\vv-\uu)\Big]} \cdot\nn\\
&\cdot& S_\o(\xx-\vv) S_{-\o}(\yy-\uu)\;.
\eea
If we put in this equation $\xx=\uu$ and $\yy=\vv$, we find, using also
\pref{Gom} and \pref{delta}, that
\bea\lb{ppp1}
&&\la\ps^+_{\xx,\o}\ps^-_{\xx,-\o}\ps^+_{\yy,-\o}\ps^-_{\yy,\o}\ra =
\la\ps^+_{\xx,\o}\ps^-_{\xx,-\o}\ps^+_{\yy,-\o}\ps^-_{\yy,\o}\ra_0 e^{-2\l_\io
[\D_+(\xx-\yy,0)-\D_-(\xx-\yy,0)] } \cr\cr &&{\sim \atop \raise6pt\hbox{
$\scriptstyle |\xx-\yy|\to \io$}} \hspace{.5cm} {C\over |\xx-\yy|^{2[1-\bar
a(0)(\l_\io/2\p)]}}\;.
\eea
If we put instead $\xx=\yy$ and $\uu=\vv$, we get
\bea\lb{ppp2}
&&\la\ps^+_{\xx,\o}\ps^+_{\xx,-\o}\ps^-_{\uu,-\o}\ps^-_{\uu,\o}\ra
=\la\ps^+_{\xx,\o}\ps^+_{\xx,-\o}\ps^-_{\uu,-\o}\ps^-_{\uu,\o}\ra_0 e^{2\l_\io
[\D_+(\xx-\uu,0)+\D_-(\xx-\uu,0)]} \cr\cr && {\sim \atop \raise6pt\hbox{
$\scriptstyle |\xx-\uu|\to \io$}} \hspace{.5cm} {C\over
|\xx-\uu|^{2[1+a(0)(\l_\io/2\p)]}}\;.
\eea
By using \pref{ppp1}, \pref{ppp2}, the first line of \pref{Aeps}, \pref{61bb}
and the definition \pref{xpm} of $x_\pm$, we finally get the first identity in
\pref{2}.

\section{Appendix: the anisotropic Ashkin-Teller model}

In this appendix, in order to derive \pref{3}, we briefly recall the analysis
of the anisotropic Ashkin-Teller model in [13]. The integration procedure is
similar to that described in \S\ref{sec2}, the main difference being that the
quadratic part \pref{2.29} of the interaction now contains also terms of the
form $\psi^{\e(\le h)}_{\xx,\o} \psi^{\e(\le h)}_{\xx,-\o}$. It follows, see
\pref{3} (where different definitions of the fermion fields were used) for
details, that we have to substitute the Grassmann integration
$P_{Z_h,\m_h}(d\psi^{(\le h)})$ in \pref{th1} with a new measure
$P_{Z_h,\m_h,\s_h}(d\psi^{(\le h)})$, where $\m_h$ and $\s_h$ are the constants
multiplying, respectively, the quadratic {\it mass terms}
\bea
2\sum_{\o=\pm} \psi^{(\le h)+}_{\xx,\o} \psi^{(\le h)-}_{\xx,-\o} \qquad{\rm
and}\qquad\ -2i\sum_{\e =\pm} \psi^{(\le h)\e}_{\xx,+} \psi^{(\le
h)\e}_{\xx,-}\;.
\eea
One can see that
\bea
&&|\log_\g(\m_{j-1}/ \m_j) - \h_\m(\l_{-\io})| \le C\l^2 \g^{\th j}\;, \cr\cr
&&|\log_\g(\s_{j-1}/ \s_j) - \h_\s(\l_{-\io})| \le C\l^2 \g^{\th j}\;.
\eea
Hence, since the two mass terms are clearly proportional, respectively, to the
operators $O^+$ and $O^-$, we find that
\be
\h_\m=\h_+-\h_z\;,\qquad \h_\s=\h_--\h_z\;.
\ee
It turns out that the difference of the critical temperatures scales as
$|v|^{x_T}$ where $x_T$, see (5.26) of [13] (where the indices are defined with
a different sign and the definitions of $\m_h$ and $\s_h$ are exchanged), is
given by
\be
x_T={1+\h_\m\over 1+\h_\s}\;,
\ee
which implies \pref{3}, since $\h_\m=1-x_+$ and $\h_\s=1-x_-$.

\*
\\

{\bf\0Acknowledgments} P.F. is indebited with David Brydges for stimulating his
interest in the topic with the request of a review seminar on the papers
\cite{[PS]} and \cite{[M1]}.

\end{document}